 \documentclass[final,1p,times]{elsarticle}

\usepackage[T1]{fontenc}
\usepackage[english]{babel}
\usepackage{amsfonts, amsmath, amssymb}
\usepackage{graphicx}
\usepackage{bm}
\usepackage{multirow}
\usepackage{color}
\usepackage{algorithm,algorithmic}
\usepackage{epstopdf}
\usepackage{placeins}
\usepackage{textcomp}
\usepackage{gensymb}
\usepackage{epsfig}
\usepackage{tabularx}
\usepackage{multirow}
\usepackage{subfigure}
\usepackage{lineno}
\usepackage{doi}
\usepackage{booktabs} 
\usepackage{adjustbox}

\graphicspath{{./Figures/}}

\newcommand{\ud}{~\mathrm{d}}
\newcommand{\NABLA}{\nabla}

\newcommand{\averg}[1]{\{ #1\}}

\newcommand{\Pol}{\mathbb{P}}

\newcommand{\mbf}{\mathbf}
\newcommand{\sbf}[1]{\boldsymbol{#1}}

\newcommand{\Fvec}{{\vec{\bf F}}}
\newcommand{\Fvecn}{{\widehat{\bf F}}}

\newcommand{\hatFvec}{\widehat{\bf F}}

\newcommand{\ie}{\emph{i.e. }}

\newcommand{\bM}{{\bf M}}
\newcommand{\bN}{{\bf N}}
\newcommand{\bR}{{\bf R}}

\newcommand{\bA}{{\bf A}}
\newcommand{\bB}{{\bf B}}
\newcommand{\bC}{{\bf C}}
\newcommand{\bD}{{\bf D}}
\newcommand{\bK}{{\bf \mathcal{K}}}
\newcommand{\bL}{{\bf \Lambda}}

\newcommand{\B}{{\bf b}}
\newcommand{\bI}{{\bf I}}
\newcommand{\bJ}{{\bf J}}

\newlength{\myVSpace}
\newcommand{\eqbydef}{\stackrel{\mathrm{def}}{=}}

\newcommand{\mathbfv}[1]{\vec{\mathbf{#1}}}

\newcommand{\uvec}{{\bf u}}

\newcommand{\vvec}{{\bf v}}
\newcommand{\velocity}{{\vec{v}}}
\newcommand{\wvec}{{\bf w}}
\newcommand{\qvec}{{\bf q}}
\newcommand{\xvec}{{\bf x}}
\newcommand{\nvec}{{\vec{n}}}

\newcommand{\Rvec}{{\bf R}}
\newcommand{\Uvec}{{\bf U}}
\newcommand{\Qvec}{{\bf Q}}
\newcommand{\Lvec}{{\bm \Lambda}}
\newcommand{\Wvec}{{\bf W}}

\journal{Journal}

\begin{document}

\begin{frontmatter}


\title{Efficient discontinuous Galerkin implementations and preconditioners for implicit unsteady compressible flow simulations}


\author[michigan,ancona]{Matteo Franciolini~\corref{cor}}\ead{mfrancio@umich.edu}

\author[michigan]{Krzysztof J. Fidkowski}

\author[ancona]{Andrea Crivellini}
 
\address[michigan]{Department of Aerospace Engineering, University of Michigan, \\1320 Beal Ave, 48109 Ann Arbor (MI), United States} 
 
\address[ancona]{Department of Industrial Engineering and Mathematical Sciences, \\ Polytechnic University of Marche, Via Brecce Bianche 12, 60131 Ancona, Italy}

\cortext[cor]{Corresponding author}

\begin{abstract}
This work presents and compares efficient implementations of high-order discontinuous Galerkin methods: a modal matrix-free discontinuous Galerkin (DG) method, a hybridizable discontinuous Galerkin (HDG) method, and a \emph{primal} formulation of HDG, applied to the implicit solution of unsteady compressible flows. The matrix-free implementation allows for a reduction of the memory footprint of the solver when dealing with implicit time-accurate discretizations. HDG reduces the number of globally-coupled degrees of freedom relative to DG, at high order, by statically condensing element-interior degrees of freedom from the system in favor of face unknowns. The \emph{primal} formulation further reduces the element-interior degrees of freedom by eliminating the gradient as a separate unknown. This paper introduces a $p$-multigrid preconditioner implementation for these discretizations and presents results for various flow problems.  Benefits of the $p$-multigrid strategy relative to simpler, less expensive, preconditioners are observed for stiff systems, such as those arising from low-Mach number flows at high-order approximation.  The $p$-multigrid preconditioner also shows excellent scalability for parallel computations.  Additional savings in both speed and memory occur with a matrix-free/reduced version of the preconditioner.


\end{abstract}


\end{frontmatter}


\section{Introduction}\label{sec:intro}
In recent years, high-order discontinuous Galerkin (DG) methods have garnered attention in the field of Computational Fluid Dynamics.  With increasing availability of high-performance computing (HPC) resources, the use of high-order methods for unsteady flow simulations has become popular. The success of these methods can be attributed to attractive dispersion and diffusion properties at high orders, ease of parallelization thanks to their compact stencil, and accuracy on unstructured meshes around complex geometries. However, the implementation of an efficient solution strategy for high-order DG methods is still a subject of active research, especially for unsteady flow problems involving the solution of the Navier--Stokes (NS) equations.

Several previous studies, see for example~\cite{bassi2007implicit, bassi2015linearly}, demonstrated that implicit schemes in the context of high-order spatial discretizations are necessary to efficiently overcome the strict stability limits of explicit time integration schemes~\cite{hesthaven2007nodal}. Implicit strategies require the solution of a large system of equations, which is typically performed with iterative solvers such as the generalized minimial residual method (GMRES). The choice of the preconditioner is a key aspect of the strategy and has been explored extensively in the literature, see for example~\cite{WANG20071994, persson2008newton, diosady2009preconditioning, birken2013preconditioning}. Among those, the use of multilevel algorithms to precondition a flexible GMRES solver~\cite{saad1993flexible} has been demonstrated as a promising choice for both compressible~\cite{WANG20071994, fidkowski2005p, shahbazi2009multigrid, diosady2009preconditioning} and incompressible flow problems~\cite{botti2017h, franciolini2018p}. Superior iterative performance compared to single-grid preconditioned solvers has been observed, as well as better parallel efficiency on distributed-memory architectures.

The application of implicit time integration strategies to DG discretizations is hindered by computational time and memory expenses associated with the assembly and storage of the residual Jacobian matrix. Although the Jacobian is sparse, the number of non-zero entries scales as $k^{2d}$, where $k$ is the approximation order and $d$ is the spatial dimension. Thus, the costs grow rapidly with approximation order, particularly in three-dimensional problems.  Motivated by this scaling, previous works~\cite{crivellini2011implicit, ceze2016development, franciolini2017efficiency, franciolini2018p} considered the possibility of a reduced matrix storage ("matrix-free") implementation of the iterative solver. This implementation avoids the allocation of the Jacobian matrix but still requires the allocation of a preconditioner operator which in some cases may still be quite large. In this context, the use of multilevel matrix-free strategies with cheap element-wise block-Jacobi preconditioners on the finest level appears to balance computational efficiency and memory considerations with iterative performance for stiff systems.  The latter is relevant to solvers applied to DG discretizations, for which the condition number scales as $\mathcal{O}(h^{-2})$~\cite{cockburn2014multigrid}, where $h$ is the mesh dimension.

The size of the DG linear system can be reduced through hybridizable discontinuous Galerkin (HDG) methods, which have been recently considered as an alternative to the standard discontinous Galerkin discretization~\cite{nguyen2009implicit, fidkowski2016hybridized}. HDG methods introduce an additional trace variable on the mesh faces but can reduce the number of globally-coupled degrees of freedom relative to DG, when a high order of polynomial approximation is employed.  The reduction occurs through a static condensation of the element-interior degrees of freedom, exploiting the block structure of the HDG Jacobian matrix. Thanks to this operation, the memory footprint of the solver scales as $k^{2(d-1)}$. Additionally, HDG methods exhibit superconvergence properties of the gradient variable in diffusion-dominated regimes. On the other hand, they increase the number of operations local to each element, both before and after the linear system solution. While several works have compared the accuracy and cost of HDG versus continuous~\cite{kirby2012cg, yakovlev2016cg} and discontinuous~\cite{woopen2014comparison, fidkowski2016hybridized} Galerkin methods, a comparison considering the efficiency of iterative solvers applied to the solution of unsteady flows is missing in this context. In fact, it is worth pointing out that, whereas HDG reduces the number of globally-coupled degrees of freedom relative to DG, at high approximation orders, its element-local operation count is non-trivial. This is particularly the case for viscous problems, in which the state gradient is approximated as a separate variable. An alternative approach is to only approximate the state, and to obtain the gradient when needed by differentiating the state. This leads to the \emph{primal} HDG formulation~\cite{dahm2017toward,devloo2018continuous}, which we also consider in this work. The advantage of \emph{primal} HDG relative to standard HDG lies mainly in the reduction of element-local operations, which translates into improved computational performance.


While for DG the use of multilevel strategies to deal with ill-conditioned systems has been previously studied, their use in HDG contexts appears not to have yet been explored, especially for unsteady flow problems. A multilevel technique has been introduced in the context of an $h$-multigrid strategy built using the trace variable projection on a continuous finite element space~\cite{cockburn2014multigrid}. In addition, the use of an algebraic multigrid method applied to a linear finite element space obtained by Galerkin projection has been proposed in the context of elliptic problems~\cite{kronbichler2016performance}. A similar idea is also considered to speed-up the iterative solution process in HDG~\cite{schutz2017hierarchical}.

The present work focuses on the comparison of implicit solution strategies in the context of unsteady flow simulations for the three aforementioned implementations, \ie DG, HDG, and \emph{primal} HDG. The comparison includes efficient preconditioning, such as $p$-multigrid, to deal with the solution of stiff linear systems arising from high-order time discretizations. In particular, an efficient algorithm to inherit the coarse space operators at a low computational cost in the context of HDG is presented for the first time. The scalability of the linear solution process is also considered and compared to standard single-grid preconditioners, such as a incomplete lower-upper factorization, ILU(0)~\cite{diosady2009preconditioning}. The efficiency of the different solution strategies and the overall memory footprint is assessed on two-dimensional laminar compressible flow problems. The results of these cases demonstrate that (i) the different discretizations attain similar error levels, (ii) the use of a multilevel strategy reduces the number of linear iterations in all cases tested, (iii) only for the DG discretizations is this advantage reflected in the CPU time, and iv) the \emph{primal} HDG and $p$-multigrid matrix-free DG solvers yield comparable solution times and memory footprints, faster than standard, single-grid preconditioners like ILU(0) applied to DG.

The paper is structured as follows. Section~\ref{sec:eqns} presents the differential equations, and Sections~\ref{sec:SPACE} and \ref{sec:TIME} show the spatial and temporal discretizations used in this work. Section~\ref{sec:MG} presents the $p$-multigrid preconditioner implementations. Section~\ref{sec:results} reports numerical experiments using different preconditioning strategies, including ILU(0), block Jacobi, and $p$-multigrid.  These experiments are performed on a range of test cases, including two-dimensional airfoils and a circular cylinder.  The methods are compared in terms of iterations, computational time, and memory footprint.

\section{Governing equations} \label{sec:eqns}

This work considers solutions of the compressible Navier--Stokes (NS) equations, which can be written as
\begin{equation}\label{eq:NS}
\begin{aligned}
      \frac{\partial \rho}{\partial t} &+
      \NABLA \cdot (\rho\velocity) = 0,\\
      \frac{\partial}{\partial t}(\rho \velocity) &+
      \NABLA \cdot(\rho \velocity \otimes \velocity + p\mathbf{I}) =
      \NABLA \cdot \bm{\tau},\\
      \frac{\partial}{\partial t} (\rho e_0) &+
      \NABLA \cdot(\rho \velocity h_{0}) =
      \NABLA \cdot \left(\velocity \cdot \bm{\tau} - \vec{q}\right),
\end{aligned}
\end{equation}
with $\velocity \in \mathbb{R}^d$ the velocity, and $d$ the number of space dimensions. The total energy $e_0$, total enthalpy $h_0$, pressure $p$, total stress tensor $\bm{\tau}$, and heat flux $\vec{q}$ are given by
\begin{center}
\begin{tabular}{ l c r }
  $e_{0} = e + (\velocity \cdot \velocity)/2$, & $h_{0} = e_0 + p/\rho$, & $p = (\gamma-1) \rho e$,  \\
  $\bm{\tau} = 2 \mu \left( \mathbf{S} - \frac{1}{3} \left(\NABLA \cdot \velocity\right) \mathbf{I} \right)$, & $\vec{q} = -\frac{\mu}{\rm{Pr}} \NABLA h$, &  $\mathbf{S} = \frac{1}{2}\left(\NABLA\velocity + (\NABLA\velocity)^T \right)$. 
\end{tabular}
\end{center}
Here $e$ is the internal energy, $h$ is the enthalpy, $\gamma=c_p/c_v$ is the ratio of gas specific heats, $\mu$ is the viscosity, $\rm{Pr}$ is the molecular Prandtl number, and $\mathbf{S}$ is the mean strain-rate tensor. The space and time discretizations are outlined in the following sections.

\section{Spatial discretization}\label{sec:SPACE}
\begin{figure}[htbp!]
\centering
   \subfigure[ DG ] {\includegraphics[width=0.35\linewidth]{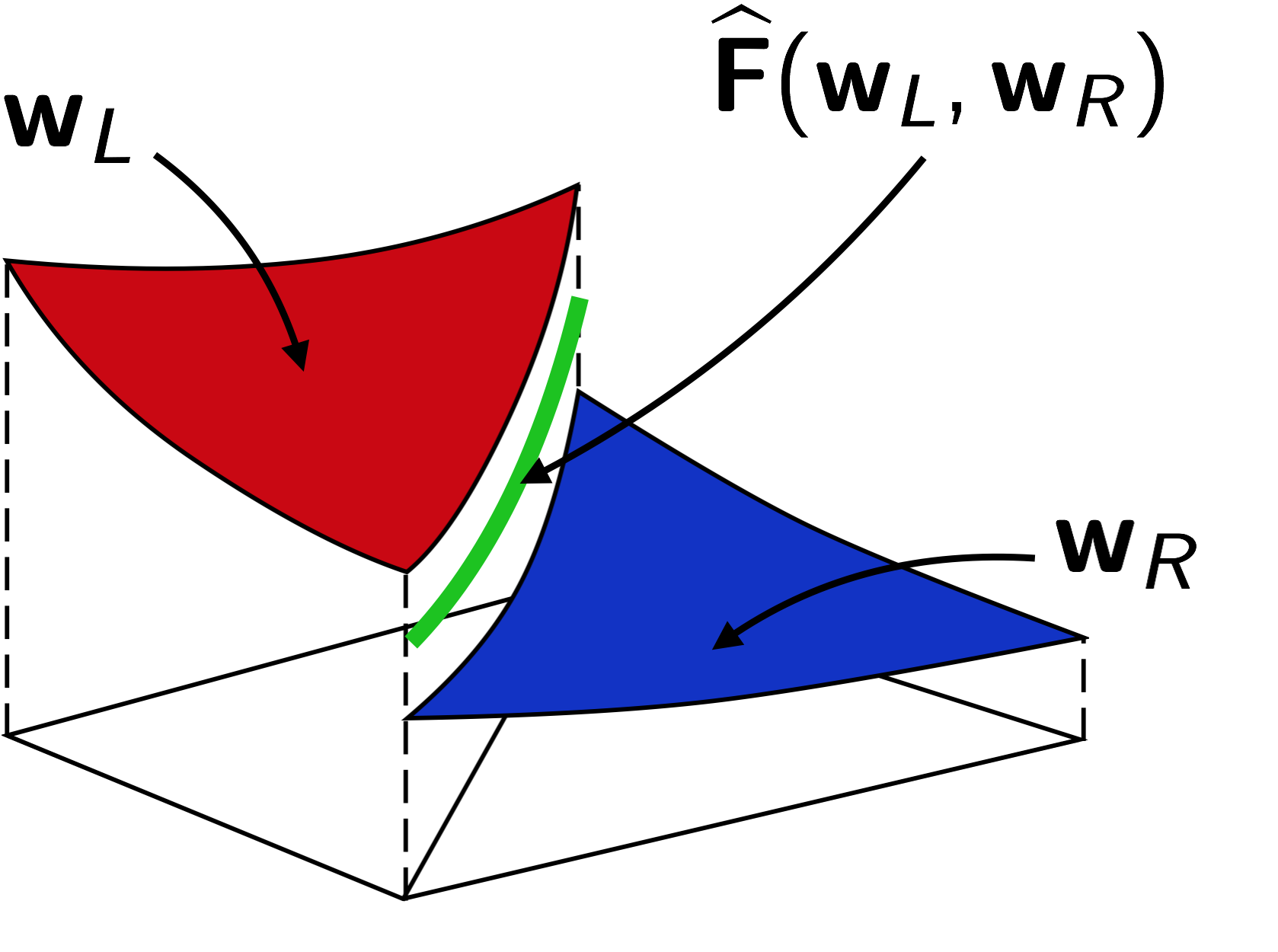}\label{fig:DGcmp}}
   \qquad
   \subfigure[ HDG ] {\includegraphics[width=0.35\linewidth]{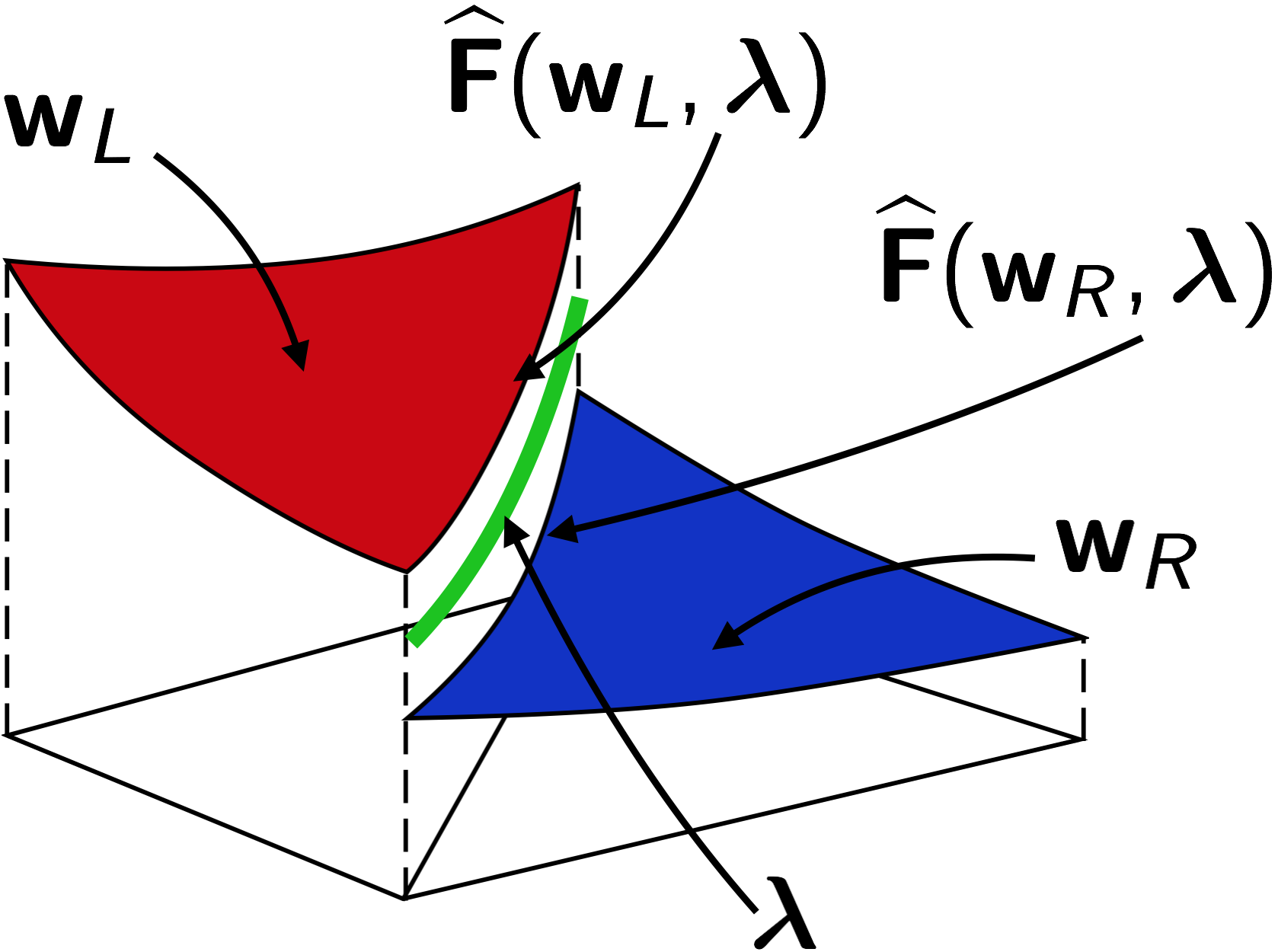}\label{fig:HDGcmp}}
\caption{Schematic comparison of the solution approximation and flux evaluation in DG versus HDG.}
\label{fig:HDGvsDG}
\end{figure}
Three versions of a modal discontinuous Galerkin (DG) finite element method are considered in this work. The first one is a standard DG implementation, which employs basis functions defined in the reference element space. The second one, commonly referred to as the hybridizable discontinuous Galerkin (HDG) method, introduces an additional set of variables defined on the mesh element interfaces to reduce the globally coupled degrees of freedom compared to DG, as shown in Figure~\ref{fig:HDGvsDG}. This implementation explicitly uses a mixed form for the gradient states (also known as the \emph{dual variable}), \emph{i.e.} the gradients are used as an additional element-wise variable, and increase the accuracy of the gradient evaluation. A third implementation, \emph{primal} HDG~\cite{dahm2017toward}, reduces the computational costs of the solver by removing the dual variable from the equations. This approach is desirable when an increased accuracy of the gradient variable is not strictly necessary or achievable for the problem. In all cases, two types of basis functions in the reference space have been considered: polynomial functions of maximum degree equal to $k$ as well as tensor product functions of degree $k$ in each dimension. In this work the former is employed within triangular grids, and the number of degrees of freedom per equation per element is $n_v^{\ell}=\prod_{i=1}^d{(k+i)/i}$. The latter approach is used for quadrilateral mesh elements, and $n_v=(k+1)^d$.

In all cases, the discretization is based on an approximation $\Omega_h$ of the domain $\Omega$ and a triangulation $\mathcal{T}_h=\{K\}$ of $\Omega_h$ made by a set of $n_e$ non-overlapping elements, denoted by $K$. Here, $\mathcal{F}_h^{i}$ stands for the set of internal element faces, $\mathcal{F}_h^{b}$ the set of boundary element faces and $\mathcal{F}_h = \mathcal{F}_h^{i} \cup \mathcal{F}_h^{b}$ their union. We also define 
\begin{equation}
  \Gamma_h^{i} = \bigcup_{F\in\mathcal{F}_h^{i}} F,
  \qquad
  \Gamma_h^{b} = \bigcup_{F\in\mathcal{F}_h^{b}} F,
  \qquad
  \Gamma_h = \Gamma_h^{i} \cup \Gamma_h^{b}.
\end{equation}
where $F$ denotes a generic mesh element face. Following Brezzi et al.~\cite{Arnold.Brezzi.ea:2002}, we also introduce the average trace operator, which on a generic internal face $F\in\mathcal{F}_h^{i}$ is defined as
$\averg{\cdot} \eqbydef \frac{(\cdot)^{+} + (\cdot)^{-}}{2}$, where $(\cdot)$ denotes a generic scalar or vector quantity. This definitions can be suitably extended to domain boundary faces by accounting for the weak imposition of boundary conditions.

\subsection{Discontinuous Galerkin}
In compact form, the system~\eqref{eq:NS} can be expressed as 
\begin{equation}\label{eq:PDE}
\dfrac{\partial \mbf{u}}{\partial t} 
+ \NABLA \cdot \mathbfv{F}_c(\mathbf{u})+\NABLA \cdot \mathbfv{F}_v(\mathbf{u}, \NABLA \mathbf{u})=\mathbf{0},
\end{equation}
where $\uvec \in \mathbb{R}^m$ is the vector of conservative variables and $\Fvec_c, \Fvec_v \in \mathbb{R}^{m{\times}d}$ are the inviscid and viscous fluxes, with $m$ the number of equations and $d$ the number of space dimensions. Note that the diffusive fluxes are connected to the gradients of $\uvec$ via the functional relation $\Fvec_d=-\mathbf{K}\NABLA \uvec$, with $\mathbf{K}$ the diffusivity tensor.

The state vector is approximated by a polynomial expansion with no continuity constraints imposed between adjacent elements, \ie $\uvec_h \in [\mathcal{V}_h]^{m}$ where 
\begin{equation}\label{eq:BSpace}
\mathcal{V}_h=\left\{\phi_h\in L_2(\Omega) : \phi_h |_{K} \in \mathbb{P}_k, \forall K \in \mathcal{T}_h\right\},
\end{equation} 
and $k$ is the order of polynomial approximation. The weak form of~\eqref{eq:PDE} follows from multiplying the PDE by the set of test functions in the same approximation space, integrating by parts, and coupling elements via numerical fluxes. The variational formulation for each element $K\in\mathcal{T}_h$ reads: find $\uvec_h \in [\mathcal{V}_h]^{m}$ such that
\begin{equation} \label{eq:dg_ns}
\begin{split}
    \int_K  \wvec_h \cdot \frac{\partial \uvec_h}{\partial t} \ud K
    -\int_{K} \NABLA_h \wvec_h : \Fvec(\uvec_h,\NABLA_h \uvec_h) \ud K \\
    + \int_{\partial K} \wvec_h^{+} \cdot \Fvecn(\uvec_h^{\pm},\NABLA_h \uvec_h^{\pm}, \nvec^+)
    \ud \sigma 
    - \int_{\partial K} \NABLA \wvec_h^+ : (\mathbf{K}^+ \cdot \nvec^+)(\uvec_h^+-\hat{\uvec}_h) \ud \sigma = 0,
\end{split}
\end{equation}
for all $\wvec_h \in [\mathcal{V}_h]^{m}$. Here, $\Fvec$ is the sum of the inviscid and viscous flux functions, while $\Fvecn$ is the numerical flux and $\hat{\uvec}_h=(\uvec_h^++\uvec_h^-)/2$. Note that the quantities $(\cdot)^+$ and $(\cdot)^-$ denote element interior and element neighbor quantities, respectively.

Uniqueness and local conservation of the solution are achieved by the use of proper numerical interface fluxes. The Roe~\cite{roe1981approximate} approximate Riemann solver is employed for the inviscid part $\hatFvec_c$, while the second form of Bassi and Rebay (BR2)~\cite{BRMPS-etc97} is employed for the viscous part, $\hatFvec_v$. Following BR2, the numerical viscous flux is given by
\begin{equation} \label{eq:visc_flux}
  \hatFvec_v\left(\uvec_h^\pm, \NABLA_h\uvec_h^\pm, \nvec^+ \right)
   \eqbydef
   \averg{\Fvec_v\left(\uvec_h, \NABLA_h\uvec_h \right)} \cdot \nvec + \eta_F \averg{\vec{\sbf \delta}_F(\uvec_h^{+}-\uvec_h^{-})} \cdot \nvec^+,
\end{equation}
where, according to~\cite{bmmpr-nmpde,Arnold.Brezzi.ea:2002}, the penalty factor $\eta_F$ must be greater than the number of faces of the elements. The auxiliary variable $\vec{\sbf \delta}$ is determined from the jump of $\uvec_h$, via the solution of the following auxiliary problem:
\begin{equation}\label{eq:re}
  \int_{K} \vec{\sbf \tau}_h^+ : \vec{\sbf \delta}_F^+ \ud K =
  \frac{1}{2} \int_{F} \vec{\sbf \tau}_h^+ : \left(\mathbf{K}^+ \cdot \nvec^+\right) (\uvec_h^{+}-\uvec_h^{-})\ud\sigma,
  \quad
  \forall{\sbf \tau}_h \in \left[\mathcal{V}_h\right]^{d{\times}m}.
 \end{equation}

At the boundary of the domain, the numerical flux function appearing in equation~(\ref{eq:dg_ns}) must be consistent with the boundary conditions of the problem. In practice, this is accomplished by properly defining a boundary state which accounts for the boundary data and, together with the internal state, allows for the computation of the numerical fluxes and the lifting operator on the portion $\Gamma_h^{b}$ of the boundary $\Gamma_h$, see~\cite{BRMPS-etc97,BR-jcpeu97}.

A system of ordinary differential equations for the degrees of freedom (DoFs) arising from Equation~\eqref{eq:dg_ns} can be compactly written in the form
\begin{equation}
\bM \dfrac{d\Wvec}{dt} + \bR(\Wvec)=\mathbf{0},
\end{equation}
where $\bM$ is the block-diagonal spatial mass matrix, $\Wvec$ is the vector of the DoFs of the problem, and $\bR$ is the spatial residual vector.

\subsection{Hybridizable discontinuous Galerkin}\label{sec:HDG}

\subsubsection{Mixed form}\label{sec:mHDG}
The second spatial discretization considered in this work is the HDG method in mixed form ($m$HDG), see~\cite{fidkowski2016hybridized}. A system of first-order partial differential equations can be obtained from~\eqref{eq:NS} by introducing $\vec{\mathbf{q}}\in \mathbb{R}^{m{\times}d}$,
\begin{equation}\label{eq:PDE1o}
\begin{split}
\vec{\mathbf{q}} - \nabla \uvec = \vec{\mathbf{0}}, &\\
\dfrac{\partial \mathbf{u}}{\partial t} + \NABLA \cdot \mathbf{F}_c(\mathbf{u})+\NABLA \cdot \mathbf{F}_v(\mathbf{u}, \vec{\mathbf{q}})=\mathbf{0}. &
\end{split}
\end{equation}
The HDG discretization approximates the state and gradient variables as $\uvec_h \in [\mathcal{V}_h]^m$ and $\vec{\mathbf{q}}_h \in [\mathcal{V}_h]^{m{\times}d}$, with $\mathcal{V}_h$ defined in \eqref{eq:BSpace}. An additional trace variable, $\bm{\lambda}_h \in [\mathcal{M}_h]^{m}$, is defined on the faces, using the space
\begin{equation}
\mathcal{M}_h=\left\{\mu \in L_2(\mathcal{F}_h^i) : \mu |_{F} \in \mathbb{P}_k, \forall F \in \mathcal{F}_h^i\right\},
\end{equation}
where $\mathbb{P}_k$ is the space of polynomials of order $k$ on face $F$. We remark that the trace variable is defined on the internal faces only, while a properly defined boundary value is used for the flux computation on $\mathcal{F}_h^b$.

The weak form is obtained by weighting the equations in~\eqref{eq:PDE1o} with appropriate test functions, integrating by parts, and using the interface variable $\bm{\lambda}_h$ for the face state. Consistent and stable numerical fluxes are required at the mesh element interfaces. The variational formulation reads: find $\uvec_h\in[\mathcal{V}_h]^m$, $\vec{\qvec}_h\in[\mathcal{V}_h]^{m{\times}d}$, $\bm{\lambda}_h\in[\mathcal{M}_h]^m$, such that
\begin{equation} \label{eq:GradHDG}
\begin{split}
    \int_{K} \vec{\vvec}_h : \vec{\qvec}_h \ud K
    +\int_{K} (\NABLA_h \cdot \vvec_h) \cdot \uvec_h \ud K
    -\int_{\partial K} (\vec{\vvec}_h^+ \cdot\nvec^+) \cdot \bm{\lambda}_h \ud \sigma = \mathbf{0},
\end{split}
\end{equation}
\begin{equation} \label{eq:NSHDG}
\begin{split}
    \int_K  \wvec_h \cdot \frac{\partial \uvec_h}{\partial t} \ud K 
    -\int_{K} \NABLA_h \wvec_h : \Fvec(\uvec_h,\vec{\qvec}_h) \ud K
    + \int_{\partial K} \wvec_h^{+} \cdot  \Fvecn(\uvec_h^+,\vec{\qvec}_h^+,\bm{\lambda}_h, \nvec^+) \ud \sigma  = \mathbf{0},
\end{split}
\end{equation}
\begin{equation} \label{eq:FluxHDG}
\begin{split}
\int_{\partial K} \bm{\mu}_h \cdot \left\{ \Fvecn(\uvec_h^{+},\vec{\qvec}_h^{+},\bm{\lambda}_h, \nvec^+) + \Fvecn(\uvec_h^{-},\vec{\qvec}_h^{-},\bm{\lambda}_h, \nvec^-) \right\} \ud \sigma = \mathbf{0},
\end{split}
\end{equation}
for all $\wvec_h\in[\mathcal{V}_h]^m$, $\vec{\vvec}_h\in[\mathcal{V}_h]^{m{\times}d}$, $\bm{\mu}_h\in[\mathcal{M}_h]^m$. The third equation, which weakly imposes flux continuity across interior faces, is required to close the system. We remark that, when using the mixed form, the same theoretical convergence rate is observed for the state variable $\uvec_h$ and the gradient variable $\vec{\qvec}_h$.  In diffusion-dominated regimes, this allows for a local post-processing of the state to a higher order~\cite{nguyen2009implicit}.

In HDG, the numerical flux function $\Fvecn$, which is the sum of the inviscid and viscous fluxes, is defined as
\begin{equation}
\Fvecn(\uvec_h,\vec{\qvec}_h,\bm{\lambda}_h, \vec{n}) = \Fvec(\bm{\lambda}_h,\vec{\qvec}_h) \cdot \nvec + \bm{\tau}(\bm{\lambda}_h,\uvec_h, \nvec),
\end{equation}
where $\bm{\tau}=\bm{\tau}_c+\bm{\tau}_v$ is a stabilization term for both the inviscid and viscous parts of the flux. In this work $\bm{\tau}_c$ is chosen in a Roe-like fashion as
\begin{equation}
\bm{\tau}_c = \left| \Fvec_c'(\bm{\lambda}_h) \cdot \vec{n} \right|(\uvec_h-\bm{\lambda}_h),
\end{equation}
while the viscous stabilization term $\bm{\tau}_v$ is based on the BR2 scheme,
\begin{equation}
\bm{\tau}_v=\eta_F \vec{\sbf \delta}_F(\uvec_h-\bm{\lambda}_h) \cdot \vec{n},
\end{equation}
with $\vec{\sbf \delta}_F$ the lifting operator applied to the jump $(\uvec_h-\bm{\lambda}_h)$, and $\eta_F$ the stabilization factor.

Similarly to DG, at the boundary of the domain, the numerical flux function is made consistent with the boundary conditions of the problem through the definition of a boundary state which accounts for the boundary data and, together with the internal state, allows for the computation of numerical fluxes and the lifting operator on the portion $\Gamma_h^{b}$ of the boundary $\Gamma_h$.

Defining $\bR^{Q}$, $\bR^{U}$ and $\bR^{\Lambda}$ as the residual vectors arising from Equations~\eqref{eq:GradHDG}, \eqref{eq:NSHDG} and \eqref{eq:FluxHDG}, the discretized system of nonlinear equations can be written as
\begin{equation}\label{eq:HDGSys}
\begin{split}
\bR^{Q}=\mathbf{0},&\\
\bM^U \dfrac{d\Uvec}{dt} + \bR^{U}=\mathbf{0}, & \\
\bR^{\Lambda}=\mathbf{0}. &
\end{split}
\end{equation}
where $\bM^U$ is the element-based mass matrix. The compact form of~\eqref{eq:HDGSys} can be written using the solution vector of the discrete unknowns, $\Wvec=[\mathbf{Q}; \Uvec; \mathbf{\Lambda}]$, and the concatenated vector of residuals $\bR=[\bR^{Q}; \bR^{U}; \bR^{\Lambda}]$,
\begin{equation}\label{eq:COMPACT}
\bM \dfrac{d\Wvec}{dt} + \bR(\Wvec)=\mathbf{0},
\end{equation}
where the matrix $\bM$ is given by
\begin{equation}
\bM = \left[
  {\begin{array}{ccc}
   \mathbf{0} & \mathbf{0} & \mathbf{0} \\
   \mathbf{0} & \bM^{U} & \mathbf{0} \\
   \mathbf{0} & \mathbf{0} & \mathbf{0}
  \end{array}}
\right].
\end{equation}

\subsubsection{Primal form}\label{sec:pHDG}
A variant of the mixed hybridizable discontinuous Galerkin method presented in Section~\ref{sec:mHDG} is the \emph{primal} HDG ($p$HDG) method and follows the work in~\cite{dahm2017toward, devloo2018continuous}. In $p$HDG, the dual variable is eliminated by introducing the definition of the gradient in~\eqref{eq:NSHDG}.

The $p$HDG discretization approximates the variable  $\uvec_h \in [\mathcal{V}_h]^m$, with $\mathcal{V}_h$ defined in Equation~\eqref{eq:BSpace}. The trace variable $\bm{\lambda}_h \in [\mathcal{M}_h]^{m}$ is still employed for hybridization, and the variational formulation reads: find $\uvec_h\in[\mathcal{V}_h]^m$, $\bm{\lambda}_h\in[\mathcal{M}_h]^m$ such that
\begin{equation} \label{eq:NSpHDG}
\begin{split}
    \int_K  \wvec_h \cdot \frac{\partial \uvec_h}{\partial t} \ud K
    -\int_{K} \NABLA_h \wvec_h : \Fvec(\uvec_h,\NABLA_h \uvec_h) \ud K & \\
    + \int_{\partial K} \wvec_h^{+} \cdot \Fvecn(\uvec_h^+,\NABLA_h \uvec_h^+,\bm{\lambda}_h, \nvec^+) \ud \sigma
    - \int_{\partial K} \NABLA \wvec_h^+ : (\mathbf{K}^+ \cdot \nvec^+)(\uvec_h^+-\hat{\uvec}_h) \ud \sigma & = \mathbf{0},
\end{split}
\end{equation}
\begin{equation} \label{eq:FluxpHDG}
\begin{split}
\int_{\partial K} \bm{\mu}_h \cdot \left\{ \Fvecn(\uvec_h^{+},\vec{\qvec}_h^{+},\bm{\lambda}_h, \nvec^+) + \Fvecn(\uvec_h^{-},\vec{\qvec}_h^{-},\bm{\lambda}_h, \nvec^-) \right\} \ud \sigma = \mathbf{0},
\end{split}
\end{equation}
for all $\wvec_h\in[\mathcal{V}_h]^m$, $\bm{\mu}_h\in[\mathcal{M}_h]^m$. We note that \eqref{eq:NSpHDG}--\eqref{eq:FluxpHDG} are not obtained by just substituting $\vec{\qvec}_h = \NABLA \uvec_h$ from \eqref{eq:GradHDG}--\eqref{eq:FluxHDG}. In fact the fourth term of \eqref{eq:NSpHDG} arises from the elimination of the variable $\vec{\qvec}_h$. This term ensures symmetry and adjoint-consistency of the \emph{primal} HDG discretization. This type of discretization, involving a smaller number of element-wise degrees of freedom than mixed HDG, does not suffer significantly from overhead costs of dealing with the gradients: eliminating the dual variable and adding the adjoint-consistency term typically results in a faster solver. We note that in this case, the gradients are of one order lower accuracy than the state variable $\uvec_h$. Numerical flux functions, stabilizing terms, and boundary condition enforcement are defined in the same manner as in mixed HDG.

Defining $\bR^{U}$ and $\bR^{\Lambda}$ as the residuals vectors arising from \eqref{eq:NSpHDG}--\eqref{eq:FluxpHDG}, the ODE system of equations can be written as
\begin{equation}\label{eq:pHDGSys}
\begin{split}
\bM^U \dfrac{d\Uvec}{dt} + \bR^{U}=\mathbf{0}, & \\
\bR^{\Lambda}=\mathbf{0}, &
\end{split}
\end{equation}
where $\bM^U$ is the elemental mass matrix. Therefore, the compact form of~\eqref{eq:pHDGSys} is written using the solution vector of discrete unknowns, $\Wvec=[\Uvec; \mathbf{\Lambda}]$, and the concatenated vector of residuals, $\bR=[\bR^{U}; \bR^{\Lambda}]$,
\begin{equation}\label{eq:COMPACTHDG}
\bM \dfrac{d\Wvec}{dt} + \bR(\Wvec)=\mathbf{0},
\end{equation}
where the matrix $\bM$ is given by
\begin{equation}
\bM = \left[
  {\begin{array}{cc}
\bM^{U} & \mathbf{0} \\
\mathbf{0} & \mathbf{0}
  \end{array}}
\right].
\end{equation}

\section{Temporal discretization} \label{sec:TIME}
The temporal discretization used in this work is an explicit-first stage, singly-diagonal-implicit Runge--Kutta (ESDIRK) scheme. The general formulation of the scheme for \eqref{eq:COMPACT} is
\begin{equation}
\begin{split}
\bM \Wvec^{i} = \bM \Wvec^n - \Delta t \sum_{j=1}^i{a_{ij}\bR(\Wvec^j)}, &\\
\Wvec^{n+1} = \Wvec^n + \Delta t\sum_{i=1}^s{\beta_i \Wvec^i}, &
\end{split}
\end{equation}
for $i=1,...,s$ where $s$ is the number of stages, $a_{ij}$ and $b_{i}$ are the coefficients of the scheme, and $n$ is the time index. Within each stage, the solution of a non-linear system is required. This is performed by the Newton-Krylov method, which requires the solution of a sequence of linear systems within each stage. In this regard, the $k$\textsuperscript{th} Newton--Krylov iteration assumes the form
\begin{equation}\label{eq:NEWTON}
\left(\dfrac{\bM}{a_{ii}\Delta t}+\dfrac{\partial\bR}{\partial \Wvec}\right)(\Wvec_{k+1}^i - \Wvec_{k}^i)=-\dfrac{\bM}{a_{ii}\Delta t}(\Wvec_k^i-\Wvec^n)-\sum_{j=1}^{i-1}\dfrac{a_{ij}}{a_{ii}}\bR(\Wvec^j)-\bR(\Wvec_k^i),
\end{equation}
with $i=1,...,s$. In this work the third-order ESDIRK3 scheme~\cite{bijl2002implicit} is employed. The method involve three non-linear solutions, following an explicit first stage.

\section{Linear system solution}
The linear system of ODEs can be solved numerically using iterative solvers. To this end, we apply the generalized minimal residual (GMRES) method to the system in~\eqref{eq:NEWTON}. The solution process differs between standard DG and HDG, as the latter discretization takes advantage of the introduction of face unknowns in order to reduce the size of the matrix to be allocated.  This section provides details of the implementation.

\subsection{Discontinuous Galerkin discretization}

The linear system arising from a DG discretizations takes the following general form 
\begin{equation}\label{eq:ODEDG}
\bK \mathbf{x} + \mathbf{b}=\mathbf{0},
\end{equation}
where $\bK = (\bM/(a_{ii}\Delta t)+ \partial \bR/\partial \Wvec)$ is the iteration matrix, $\mathbf{x}=\Delta\Wvec$ is the vector of degrees of freedom updates, and $\mathbf{b}$ is the right-hand side. The GMRES implementation can follow two approaches. The first one is denoted as matrix-based (MB) and consists of computing and storing the iteration matrix explicitly to perform matrix-vector products as needed within the iterative solution. A second, matrix-free (MF) approach takes advantage of the structure of the matrix vector products, which can be approximated using the matrix-free formula
\begin{equation}\label{eq:DG-mf}
\left(\dfrac{\bM}{a_{ii}\Delta t}+\dfrac{\partial\bR}{\partial \Wvec}\right) \Delta \Wvec  \approx \dfrac{\bM}{a_{ii}\Delta t} \Delta \Wvec + \dfrac{\bR(\Wvec+h\Delta\Wvec)-\bR(\Wvec)}{h},
\end{equation}
with
\begin{equation}
h=\varepsilon\dfrac{\sqrt{1+\|\Delta \Wvec\|}}{\|\Wvec\|}
\end{equation}
and $\varepsilon \approx 10^{-7}$. The latter approach offers several advantages over the former, as the Jacobian matrix is no longer required to maintain the temporal accuracy of the solution. The GMRES solver still requires a preconditioning matrix, which generally needs to be stored. However, this matrix can be approximated or frozen for a certain number of iterations without losing the formal order of accuracy of the time integration scheme, thus providing an improvement in computational efficiency. For example, when using a block-Jacobi preconditioning approach, the memory footprint required for the Jacobian matrix can be one order of magnitude lower, as only the memory for the on-diagonal blocks needs to be allocated. The additional residual evaluation, which are required in~\eqref{eq:DG-mf}, have a computational cost similar to a matrix-vector product for high-order polynomials. Further details can be found in previous work~\cite{franciolini2017efficiency, crivellini2011implicit}.

\subsection{Hybridizable discontinuous Galerkin discretization}

\subsubsection{Mixed form}

The system in \eqref{eq:NEWTON} can be conveniently arranged using the definition of element-interior and face DoFs. To this end, considering first the mixed form of the HDG discretization, it is convenient to define the following elemental block matrices at stage $i$ of the Newton-Krylov method
\begin{equation}
\begin{array}{lll}
\bA^{QQ}=\dfrac{\partial \Rvec^Q}{\partial \Qvec}, & \bA^{QU}=\dfrac{\partial \Rvec^Q}{\partial \Uvec}, & \mathbf{B}^{Q\Lambda}= \dfrac{\partial \Rvec^Q}{\partial \Lvec}, \\[2.5mm]
\bA^{UQ}=\dfrac{\partial \Rvec^U}{\partial \Qvec}, & \bA^{UU}=\dfrac{\bM^U}{a_{ii}\Delta t}+\dfrac{\partial \Rvec^U}{\partial \Uvec}, & \mathbf{B}^{U\Lambda}=\dfrac{\partial \Rvec^U}{\partial \Lvec}, \\[2.5mm]
\mathbf{C}^{\Lambda Q}=\dfrac{\partial \Rvec^{\Lambda}}{\partial \Qvec}, & \mathbf{C}^{\Lambda U}=\dfrac{\partial \Rvec^{\Lambda}}{\partial \Uvec}, & \mathbf{D}=\dfrac{\partial \Rvec^{\Lambda}}{\partial \Lvec},
\end{array}
\end{equation}
while the right hand side of Eq.~\eqref{eq:NEWTON} is obtained as
\begin{equation}
{\begin{array}{l}
   \mathbf{f}^{Q}=-\displaystyle{\sum_{j=1}^{i-1}}\dfrac{a_{ij}}{a_{ii}}\bR^Q(\Qvec^j,\Uvec^j,\Lvec^j)-\bR^Q(\Qvec_k^i,\Uvec_k^i,\Lvec_k^i), \\
   \mathbf{f}^{U}=-\dfrac{\bM^U}{a_{ii}\Delta t}\left(\Uvec_k^i-\Uvec^n\right)-\displaystyle{\sum_{j=1}^{i-1}}\dfrac{a_{ij}}{a_{ii}}\bR^U(\Qvec^j,\Uvec^j,\Lvec^j)-\bR^U(\Qvec_k^i,\Uvec_k^i,\Lvec_k^i), \\
   \mathbf{g}=-\displaystyle{\sum_{j=1}^{i-1}}\dfrac{a_{ij}}{a_{ii}}\bR^{\Lambda}(\Qvec^j,\Uvec^j,\Lvec^j)-\bR^{\Lambda}(\Qvec_k^i,\Uvec_k^i,\Lvec_k^i).
  \end{array}}
\end{equation}
The full system of equations can be therefore written in the compact form as
\begin{equation}\label{eq:HDGMat}
\left[
  {\begin{array}{lll}
   \bA^{QQ} & \bA^{QU} & \mathbf{B}^{Q\Lambda} \\
   \bA^{UQ} & \bA^{UU} & \mathbf{B}^{U\Lambda} \\
   \mathbf{C}^{\Lambda Q} & \mathbf{C}^{\Lambda U} & \mathbf{D} 
  \end{array}}
\right]\left(
  {\begin{array}{c}
   \Delta \mathbf{Q} \\
   \Delta \mathbf{U}  \\
   \Delta \mathbf{\Lambda} 
  \end{array}}
\right)+\left(
  {\begin{array}{c}
   \mathbf{f}^{Q} \\
   \mathbf{f}^{U} \\
   \mathbf{g} 
  \end{array}}
\right)=\mathbf{0}.
\end{equation}

\subsubsection{Primal form}
In the \emph{primal} formulation, the elemental block matrices related to the gradient variables are no longer present in the linear system. Moreover, the right-hand side can be evaluated via the following equations
\begin{equation}
{\begin{array}{l}
   \mathbf{f}^{U}=-\dfrac{\bM^U}{a_{ii}\Delta t}\left(\Uvec_k^i-\Uvec^n\right)-\displaystyle{\sum_{j=1}^{i-1}}\dfrac{a_{ij}}{a_{ii}}\bR^U(\Uvec^j,\Lvec^j)-\bR^U(\Uvec_k^i,\Lvec_k^i), \\
   \mathbf{g}=-\displaystyle{\sum_{j=1}^{i-1}}\dfrac{a_{ij}}{a_{ii}}\bR^{\Lambda}(\Uvec^j,\Lvec^j)-\bR^{\Lambda}(\Uvec_k^i,\Lvec_k^i).
  \end{array}}
\end{equation}
that do not depend anymore on the internal DoFs $\Qvec$. The linear system for the primal HDG discretization assumes the form
\begin{equation}\label{eq:pHDGMat}
\left[
  {\begin{array}{ll}
   \bA^{UU} & \mathbf{B}^{U\Lambda} \\
   \mathbf{C}^{\Lambda U} & \mathbf{D} 
  \end{array}}
\right]\left(
  {\begin{array}{c}
   \Delta \mathbf{U}  \\
   \Delta \mathbf{\Lambda} 
  \end{array}}
\right)+\left(
  {\begin{array}{c}
   \mathbf{f}^{U} \\
   \mathbf{g} 
  \end{array}}
\right)=\mathbf{0},
\end{equation}

\subsection{Static condensation and back-solve}

Considering the block structure of the matrices appearing in~\eqref{eq:HDGMat} and \eqref{eq:pHDGMat}, the solution of the system can involve a smaller number of DoFs by statically condensing out the element-interior variables. Partitioning the matrix into element-interior and face components, $[\bA, \mathbf{B}; \mathbf{C}, \mathbf{D}]$, and similarly for the right-hand side vector, $[\mathbf{f};\mathbf{g}]$, the Schur-complement linear system reads
\begin{equation}\label{eq:HDGCOND}
\underbrace{(\bD-\bC\bA^{-1}\bB)}_{\mathcal{K}}\Delta \bL=\underbrace{(\mathbf{g}-\bC\bA^{-1}[\mathbf{f}])}_{\B},
\end{equation}
which assumes the same form as system~\eqref{eq:ODEDG} and can be solved using a GMRES algorithm. The definition of each block can be found from~\eqref{eq:HDGMat} and \eqref{eq:pHDGMat}. The static condensation is an operation that involves matrix-matrix products for the iteration matrix, as well as matrix-vector products for the right-hand side. Fortunately, the compact structure of the residual Jacobian prevents us from having to allocate global matrices for the computation of the condensed matrix, \emph{i.e.} the operations described in Eq.~\eqref{eq:HDGCOND} are local to each element. In addition, the computation of $\bA^{-1}$ can be performed in place. By doing so, we do not increase the memory footprint of the HDG implementation during the solve.

After the solution of~\eqref{eq:HDGCOND}, the interior states have to be recovered for the residual evaluation in the next time step. This operation is commonly referred to as the \emph{back solve} and assumes the following form
\begin{equation}\label{eq:HDGBACKSOLVE}
\Delta\Uvec = -\bA^{-1} \left( \mathbf{f} +\bB\Delta \bL\right).
\end{equation}
Our implementation choice of assembling the condensed matrix on-the-fly requires re-evaluation of the inverse of the matrix $\bA$ in an element-wise fashion during the back-solve. 

As a final remark for the two solvers, we point out that for both the \emph{mixed} and \emph{primal} form of HDG, memory allocation and time spent on the global solve are lower than that of a DG solver due to the smaller number of globally-coupled degrees of freedom at high orders. On the other hand, the inversion of the $\bA$ block-structured matrix of equation~\eqref{eq:HDGCOND}, although local to each element, increases the amount of element-wise operations.

\section{Multigrid preconditioning} \label{sec:MG}
The use of a $p$-multigrid strategy to precondition a flexible implementation~\cite{saad1993flexible} of GMRES is explored in the context of the spatial discretizations presented. The basic multigrid idea is to exploit iterative solvers to smooth-out high-frequency components of the error with respect to an unknown exact solution. Such iterative solvers are not sufficiently effective at damping low-frequency error components, and to this end, an iterative solution built using coarser problems can be useful to shift, via system projection, the low-frequency modes towards the upper side of the spectrum. This simple and effective strategy allows us to obtain satisfactory rates of convergence over the entire range of error frequencies.

In \emph{p}-multigrid the coarse problems are built based on lower-order discretizations with respect to the original problem of degree $k$. We consider $L$ coarse levels denoted by the index $\ell = 0,...,L$ and indicate the fine and coarse levels with $\ell=0$ and $\ell=L$, respectively. The polynomial degree of level $\ell$ is $k_{\ell}$ and the polynomial degrees of the coarse levels are chosen such that $k_{\ell} < k_{\ell-1}$, with $k_0 = k$.  These orders are used to build coarser linear systems $\bK_{\ell} \xvec_{\ell} = \mathbf{b}_{\ell}$. In order to avoid additional integration of the residuals and Jacobians on the coarse level, we employ subspace inheritance of the matrix operators assembled on the finest space to build the coarse space operators $\bK_i$ for both DG and HDG discretizations. This choice involves projections of the matrix operators and right hand sides, which are computed only once on the finest level. Compared to subspace non-inheritance, which requires the re-evaluation of the Jacobians in proper coarser-space discretizations of the problem, inheritance is cheaper in processing and memory. Although previous work has shown lower convergence rates when using such cheaper operators~\cite{antonietti2015multigrid, botti2017h, franciolini2018p}, especially in the context of elliptic problems and incompressible flows, we found these operators sufficiently efficient for our target problems involving the compressible NS equations, as will be demonstrated in the results section.

In the context of standard discontinuous Galerkin discretizations, see for example \cite{fidkowski2005p,diosady2009preconditioning,shahbazi2009multigrid,botti2017h,franciolini2018p}, the $p$-multigrid approach has been thoroughly investigated and exploited in several ways, \emph{e.g.} $h$-, $p$-, and $hp$-strategies. On the other hand, the use of $p$-multigrid for HDG has not been as widely studied. See, for example, preliminary works~\cite{cockburn2014multigrid,kronbichler2016performance,schutz2017hierarchical} related to this research area.
The definition of the restriction and prolongation operators, as well as the coarse grid operators and right hand sides, is not straightforward when considering the statically condensed system.  We will therefore first introduce the concept of subspace inheritance for a standard DG solver and then extend it to HDG.

We employ a full multigrid (FMG) $\mathcal{V}$-cycle solver, outlined in Algorithm~\ref{ALG:FMG}. The FMG cycle constructs a good initial guess for a $\mathcal{V}$-cycle iteration, which starts on the fine space. To do so, the solution is initially obtained on the coarsest level ($L$), and then prolongated to the next refined one ($L-1$). At this point, a standard $\mathcal{V}$-cycle is called, such that an improved approximation of the solution can be used for the $\mathcal{V}$-cycle at level $L-2$. This procedure is repeated until the $\mathcal{V}$-cycle on the finest level is completed. 
\begin{figure}[t!]
\begin{minipage}[c]{0.495\linewidth}
\begin{algorithm}[H]
\begin{algorithmic}[1]
\FOR{$\ell=L,0,-1$}
\IF{$\ell = L$}
	\STATE{$\mathbf{b}_{\ell} = \mathbf{I}_{0}^{\ell}\mathbf{b}_{0}$}
	\STATE {SOLVE $\bA_{\ell}{\mathbf{x}}_{\ell}^{FMG}=\mathbf{b}_{\ell}$} 
	\ELSE
	\STATE{$\mathbf{b}_{\ell} = \mathbf{I}_{0}^{\ell}{\mathbf{b}}_{0}$}
	\STATE{$\tilde{\mathbf{x}}_{\ell} = \mathbf{I}_{\ell+1}^{\ell}\mathbf{x}_{\ell+1}^{FMG}$}
	\STATE{$\mathbf{x}_{\ell}^{FMG}=MG_V(\ell,\mathbf{b}_{\ell}, \tilde{\mathbf{x}}_{\ell} )$}
	\ENDIF
\ENDFOR
\RETURN {$\mathbf{x}_{\ell}^{FMG}$}
\end{algorithmic}
\caption{$MG_{\textrm{full}}$}
\label{ALG:FMG}
\end{algorithm}
\end{minipage}
\begin{minipage}[c]{0.495\linewidth}
\begin{algorithm}[H]
\begin{algorithmic}[1]
\IF{$\ell = L$}
\STATE {SOLVE $\bA_{\ell}\overline{\mathbf{x}}_{\ell}=\mathbf{b}_{\ell}$} 
\ELSE
\STATE {$\overline{\mathbf{x}}_{\ell}$=SMOOTH($\mathbf{x}_{\ell},\bA_{\ell},\mathbf{b}_{\ell}$)} 
\STATE {$\mathbf{r}_{\ell}=\mathbf{b}_{\ell}-\bA_{\ell}\overline{\mathbf{x}}_{\ell} $}
\STATE {$\mathbf{r}_{\ell+1}=\mathbf{I}_{\ell}^{\ell+1}\mathbf{r}_{\ell}$}
\STATE {$\mathbf{e}_{\ell+1}$=$MG_V(\ell+1,\mathbf{r}_{\ell+1},\mathbf{0})$}  
\STATE {$\hat{\mathbf{x}}_{\ell}=\overline{\mathbf{x}}_{\ell}+\mathbf{I}_{\ell+1}^{\ell}\mathbf{e}_{\ell+1}$}
\STATE {$\overline{\mathbf{x}}_{\ell}$=SMOOTH($\hat{\mathbf{x}}_{\ell},\bA_{\ell},\mathbf{b}_{\ell}$)} 
\ENDIF
\RETURN $\overline{\mathbf{x}}_{\ell}$
\end{algorithmic}
\caption{$MG_{\mathcal{V}}(\ell,\mathbf{b}_{\ell}, \mathbf{x}_{\ell})$}
\label{ALG:VMG}
\end{algorithm}
\end{minipage}
\end{figure}
The single $\mathcal{V}$-cycle is outlined in Algorithm~\ref{ALG:VMG}. Starting from a level $\ell$, the solution is initially smoothed using an iterative solver (SMOOTH). The residual of the solution, $\mathbf{r}_{\ell}$, is then computed and projected to the coarser level $\ell+1$, where another $\mathcal{V}$-cycle is recursively called to obtain a coarse-grid correction $\mathbf{e}_{\ell+1}$. This quantity is prolongated to level $\ell$ and used to correct the solution to be smoothed again. When the coarsest level is reached, the problem is solved with a larger number of iterations to decrease as much as possible the solution error at a low computational cost.

In this work the smoothers consist of preconditioned GMRES solvers. Any combination of single grid preconditioners can be coupled with an iterative solver to properly operate as a smoother in a multigrid strategy.
Devising a methodology to optimally set the multigrid preconditioner is beyond the scope of the present work. However, previous work~\cite{franciolini2018p} has shown that an optimal and scalable solver can be obtained using an aggressive preconditioner on the coarsest space discretization, where the factorization of the matrix can be performed at a low computational cost, and the system has to be solved with a higher accuracy. On the other hand, cheaper operators can be used on the finest levels of the discretization, where the systems need not be solved to a high degree of accuracy.
In the following subsections, the details are given about how the matrices and vectors are restricted and prolongated between multigrid levels.

\subsection{DG subspace inheritance}\label{sec:DGInheritance}
Let us define a sequence of approximation spaces $\mathcal{V}_{\ell}\subseteq\mathcal{V}_{h}$ on the same triangulation $\mathcal{T}_h$, $\ell$ being a multigrid level, with $\mathcal{V}_{h}=\mathcal{V}_{0} \supset \mathcal{V}_{1}  \supset ... \supset \mathcal{V}_{L}$ and $L$ the number of coarse levels. Note that $\mathcal{V}_{L}$ denotes the coarsest space. In our $p$-multigrid setting, each approximation space is defined similarly to~\eqref{eq:BSpace}, but using $k_{\ell}$ polynomials, with $k_0>...>k_{\ell} > ... > k_{L}$.

In this context, the prolongation operator can be defined as $\mathcal{I}_{\ell+1}^{\ell}:\mathcal{V}_{\ell+1}\rightarrow \mathcal{V}_{\ell}$ such that
\begin{equation}
\sum_{K\in\mathcal{T}_h}\int_K{\left(\mathcal{I}_{\ell+1}^{\ell}u_{\ell+1}-u_{\ell+1}\right)dK}=0, \hspace{0.5cm} \forall u_{\ell+1} \in \mathcal{V}_{\ell+1}.
\end{equation}
Similarly, the restriction operator can be defined as the $L_2$ projection $\mathcal{I}_{\ell}^{\ell+1}:\mathcal{V}_{\ell}\rightarrow \mathcal{V}_{\ell+1}$ such that
\begin{equation}
\sum_{K\in\mathcal{T}_h}\int_K{\left(\mathcal{I}_{\ell}^{\ell+1}u_{\ell}-u_{\ell}\right)v_{\ell+1}dK}=0, \hspace{0.5cm} \forall (u_{\ell},v_{\ell+1}) \in \mathcal{V}_{\ell} \times \mathcal{V}_{\ell+1}.
\end{equation}
Such a definition can be extended to operate on vector functions $\uvec_h\in[\mathcal{V}_{\ell}]^m$ component-wise, \emph{i.e.} $\mathcal{I}_{\ell}^{\ell+1}\uvec_{h}= \left\{ \mathcal{I}_{\ell}^{\ell+1}u_{i} \right\} $.

Regarding coarse-space matrices, discrete matrix operators $\mathbf{I}_{\ell}^{\ell+1}\in\ \mathbb{R}^{p{\times}q}$, with $p=n_e n_{\textrm{v}}^{{\ell}}m$, $q=n_e n_{\textrm{v}}^{{\ell+1}} m$ and $n_v^{\ell}$ the number of DoFs on level $\ell$, have to be considered to inherit the fine-space iteration matrix $\bK_0$. This matrix can be restricted up to level $\ell$ via $\bK_{\ell}=(\bI_{0}^{\ell})\bK_0(\bI_{\ell}^{0})$. Fortunately, the explicit assembly of the operators can be avoided and the projection can be applied for each matrix block $b$ of size $(n_{v}^{\ell}m)^2$, which assumes the following form
\begin{equation}
\mathcal{K}_{\ell+1}^{b}=\bM_{{\ell+1},{\ell}}\mathcal{K}_{\ell}^{b}\left(\bM_{{\ell+1},{\ell}}\right)^T,
\end{equation}
where
\begin{equation}
\left(\bM_{{\ell+1},{\ell}}\right)=\left(\bM_{\ell+1}\right)^{-1}\int_K{\bm{\phi}^{{\ell+1}}\otimes\bm{\phi}^{{\ell}}\ud K}, \quad \quad \bM_{\ell+1}= \int_K{\bm{\phi}^{{\ell+1}}\otimes\bm{\phi}^{{\ell+1}}\ud K}.
\end{equation}
Here, $\bm{\phi}^{{\ell}}$ denotes the set of basis functions defined in element $K$ of order $k_{\ell}$. We point out that, thanks to the use of basis functions defined in a reference element, the projection operators are identical for each element and pair of polynomial orders, and they can be used in the same way for the on-diagonal and off-diagonal blocks of the iteration matrix. As the prolongation operators can be obtained from the restriction by the transpose, $\bI_{\ell+1}^{\ell}=\left(\bI_{\ell}^{\ell+1}\right)^T$, the method requires the allocation of only $L$ matrices of size $n_{v}^{{\ell+1}}{\times}n_{v}^{{\ell}}$ for each different type of element, which is inexpensive from a memory footprint viewpoint.

\subsection{HDG subspace approximate-inheritance}\label{sec:HDGInheritance}

In HDG, the globally-coupled unknowns are those related to the face DoFs, and the iteration matrix is obtained through static condensation, see Equation~\eqref{eq:HDGCOND},
which allows us to solve the system for the face unknowns only. Theoretically speaking, for element-interior degrees of freedom, the same operators of Section~\ref{sec:DGInheritance} can be employed, while those for faces degrees of freedom can be obtained through similar considerations. In this case, the sequence of approximation spaces $\mathcal{M}_{\ell}\subseteq\mathcal{M}_{h}$ is properly defined on the interior mesh element faces, $\mathcal{F}_h$, with $\ell$ a multigrid level. To this end, we define $\mathcal{M}_{h}=\mathcal{M}_{0} \supset \mathcal{M}_{1}  \supset ... \supset \mathcal{M}_{L}$. The prolongation operator is now $\mathcal{J}_{\ell+1}^{\ell}:\mathcal{M}_{\ell+1}\rightarrow \mathcal{M}_{\ell}$, defined by
\begin{equation}
\sum_{F\in\mathcal{F}_h}\int_F{\left(\mathcal{J}_{\ell+1}^{\ell}\lambda_{\ell+1}-\lambda_{\ell+1}\right)} \ud \sigma =0, \hspace{0.5cm} \forall \lambda_{\ell+1} \in \mathcal{M}_{\ell+1}.
\end{equation}
The restriction operator is defined as $\mathcal{J}_{\ell}^{\ell+1}:\mathcal{M}_{\ell}\rightarrow \mathcal{M}_{\ell+1}$ such that
\begin{equation}
\sum_{F\in\mathcal{F}_h}\int_F{\left(\mathcal{J}_{\ell}^{\ell+1}\lambda_{\ell}-\lambda_{\ell}\right)\mu_{\ell+1}} \ud \sigma =0, \hspace{0.5cm} \forall (\lambda_{\ell},\mu_{\ell+1}) \in \mathcal{M}_{\ell} \times \mathcal{M}_{\ell+1}.
\end{equation}
These definitions can also be extended to operate on vector functions $\bm{\lambda}_h\in[\mathcal{M}_{\ell}]^M$ and are assumed to act component-wise.


Applying the same subspace-inheritance idea used for DG, one obtains the coarse space condensed HDG matrix and right hand side through the application of element-interior and face DoF projections,
\begin{equation}\label{eq:HDGInh}
\bK_{\ell}=\left(
(\bJ_0^{\ell})\mathbf{D}_0(\bJ_{\ell}^0)-
\left[
(\bJ_0^{\ell})\mathbf{C}_0(\bI_{\ell}^0)
\right]
\left[
(\bI_0^{\ell})\bA_0(\bI_{\ell}^0) 
\right]^{-1}
\left[
(\bI_0^{\ell})\mathbf{B}_0(\bJ_{\ell}^0)
\right]
\right),
\end{equation}
\begin{equation}\label{eq:HDGRHSInh}
\mathbf{b}_{\ell}=\left(
(\bJ_0^{\ell})\mathbf{g}_0-
\left[
(\bJ_0^{\ell})\mathbf{C}_0(\bI_{\ell}^0) 
\right]
\left[
(\bI_0^{\ell})\bA_0(\bI_{\ell}^0)
\right]^{-1}\left[
(\bI_0^{\ell})\mathbf{f}_0
\right]\right).
\end{equation}
We point out that this operation involves the application of mixed element-interior and face degrees of freedom Galerkin projections prior to the static condensation of the system. Thus, it comes with an increased operation count compared to DG subspace inheritance. To minimize the number of operations involved in the projection, we introduce an \emph{approximate}-inherited approach where the coarse space matrices and right-hand sides are obtained by simply applying face projections to the condensed matrices and vectors on the finest space. In other words, we obtain $\bK_{\ell}$ and $\mathbf{b}_{\ell}$ through
\begin{equation}
\bK_{\ell}=
(\bJ_0^{\ell})\bK_0(\bJ_{\ell}^0),
\end{equation}
\begin{equation}
\mathbf{b}_{\ell}=
(\bJ_0^{\ell})\mathbf{b}_0.
\end{equation}
This coarse space matrix and right-hand side will in general differ from those of Equation~\eqref{eq:HDGInh} and~\eqref{eq:HDGRHSInh}, as the element-interior operator in the Schur complement term is projected differently.

Similar considerations have to be made for the residual evaluation on the coarse levels, which are required in the projection from level $\ell$ to $\ell+1$. The residual on a level $\ell$ should be computed as 
\begin{equation}
\mathbf{r}_{\ell}=\left(
\mathbf{R}_{\ell}^{\Lambda}-
\left[
(\bJ_0^{\ell})\mathbf{C}_0(\bI_{\ell}^0)
\right]
\left[
(\bI_0^{\ell})\bA_0^{QQ}(\bI_{\ell}^0) 
\right]^{-1}
\mathbf{R}_{\ell}^{QU},
\right)
\end{equation}
and this requires the evaluation of the element-interior degrees of freedom from the face unknowns. This operation would require a back-solve on the coarse levels, which increases the amount of operations within each multigrid cycle.  To reduce the operation count as much as possible we again rely on the computation of an approximate projection that is evaluated using the working variable $\Delta \mathbf{\Lambda}$ only:
\begin{equation}
\mathbf{r}_{\ell}=\bK_{\ell} \Delta \mathbf{\Lambda}_{\ell} + \mathbf{b}_{\ell}.
\end{equation}
Despite the use of those approximations compared to DG subspace inheritance, such a strategy exhibits very good performance in HDG solutions of the compressible Navier--Stokes equations, as will be demonstrated in the results section.

Similarly to DG, the global assembly of the projection matrix $\bJ_{0}^{\ell}$ for face degrees of freedom is not strictly required for HDG, since the projection is applied to each block $b$ of the matrix of size $(n_{\textrm{v}}^{\ell} m)^2$, with $n_{\textrm{v}}^{\ell}$ related to face degrees of freedom on level $\ell$. The projection of each block $b$ of the iteration matrix assumes the form
\begin{equation}
\mathcal{K}_{\ell+1}^{b}=\bN_{{\ell+1},{\ell}}\mathcal{K}_{\ell}^{b}\left(\bN_{{\ell+1},{\ell}}\right)^T,
\end{equation}
where
\begin{equation}
\left(\bN_{{\ell+1},{\ell}}\right)=\left(\bN_{\ell+1}\right)^{-1}\int_F{\bm{\mu}^{{\ell+1}}\otimes\bm{\mu}^{{\ell}}\ud \sigma}, \quad \quad \bN_{\ell+1}= \int_F{\bm{\mu}^{{\ell+1}}\otimes\bm{\mu}^{{\ell+1}}\ud \sigma}.
\end{equation}
In this case $\bm{\mu}^{{\ell}}$ denotes the set of basis functions defined in the element face $F$ of order $k_{\ell}$. Thanks to the use of basis functions defined in the reference element of the space discretization, similar considerations to those related to element-interior degrees of freedom projection hold true.

\subsection{Preconditioning options and memory footprint}

In this work we exploit single-grid preconditioners both for benchmarking and to precondition the smoothers of the multigrid strategy. Two operators will be considered in this work. The first and simplest one will be here labelled as element-wise block-Jacobi (BJ). The BJ preconditioner extracts the block-diagonal portion of the iteration matrix and factorize it, in a local-to-each element fashion, using the PLU factorization. This cheap and memory saving preconditioner becomes more effective as the time step size of the discretization decreases, \emph{i.e.} the matrix becomes more diagonally dominant and the condition number decreases. In addition, the BJ preconditioner is applied in the same way in serial and parallel computations.


The second preconditioner is the incomplete lower-upper factorization with zero-fill, ILU(0). In particular, the minimum discarded fill reordering proposed in~\cite{persson2008newton} is employed. The algorithm showed to be suited for stiff spatial discretizations, and it allows an in-place factorization~\cite{diosady2009preconditioning}. When it is applied in parallel, the ILU(0) is performed on each square, partition-wise block of the iteration matrix. In this case this preconditioner will be labelled as block-ILU(0) (BILU). As a natural downside, BILU loses the preconditioning efficiency when it is applied in parallel. A variant that compensate this effect is the Additive Schwarz method, which extends the partition-wise block of the Jacobian with a number of overlapping elements between the mesh partition. This algorithm, employed in the context of incompressible Navier--Stokes equations, increases the memory footprint of the solver when a few elements per partition is used~\cite{franciolini2018p}. We do not use such an approach in the present study of compressible flow.

\begin{figure}[b!]
 \centering
  \subfigure[Full-order basis functions]{
  \includegraphics[width=0.485\textwidth]{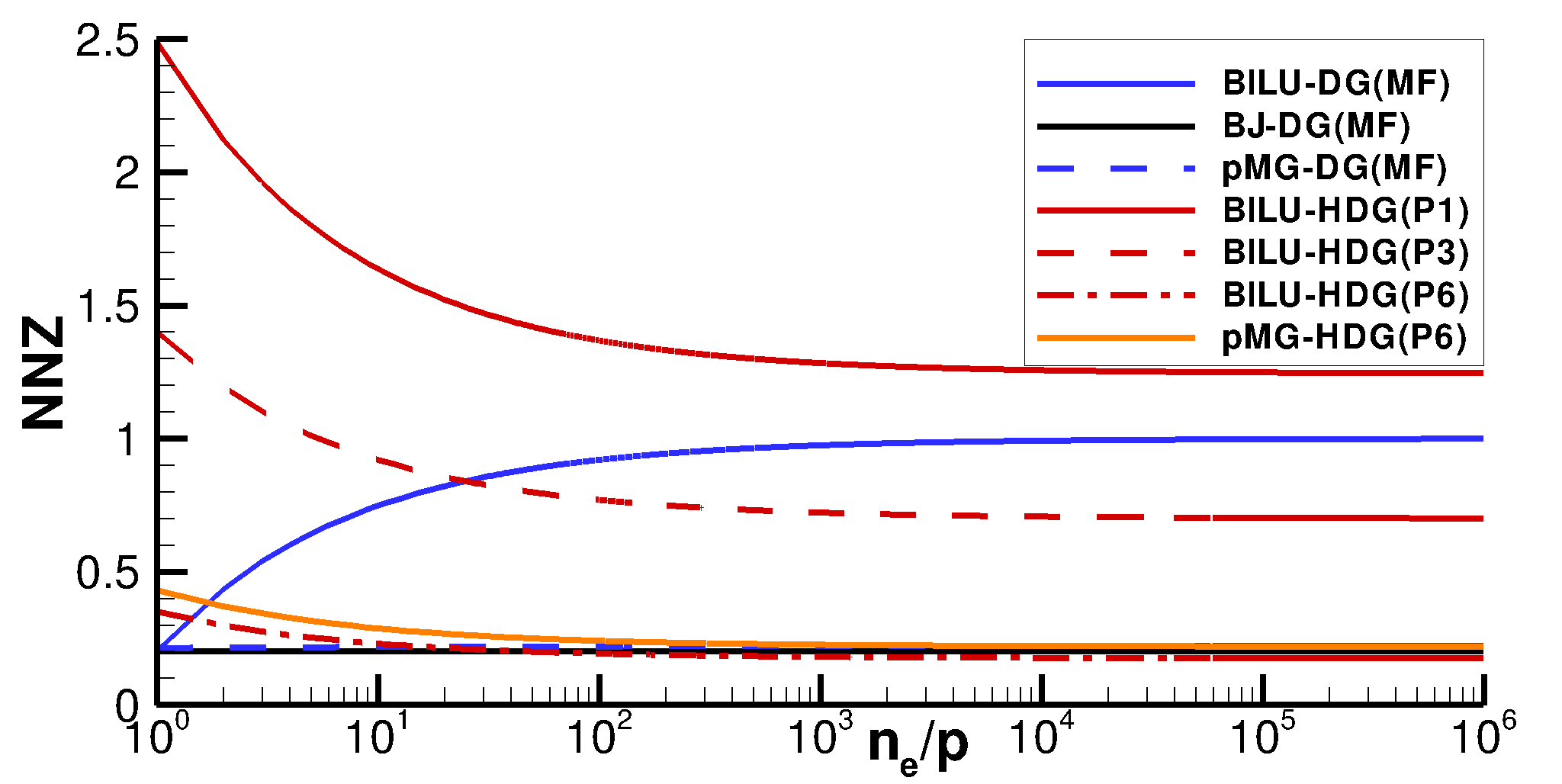}\label{fig:TriLagrangeMem}}
  \subfigure[Tensor-product basis functions]{
  \includegraphics[width=0.485\textwidth]{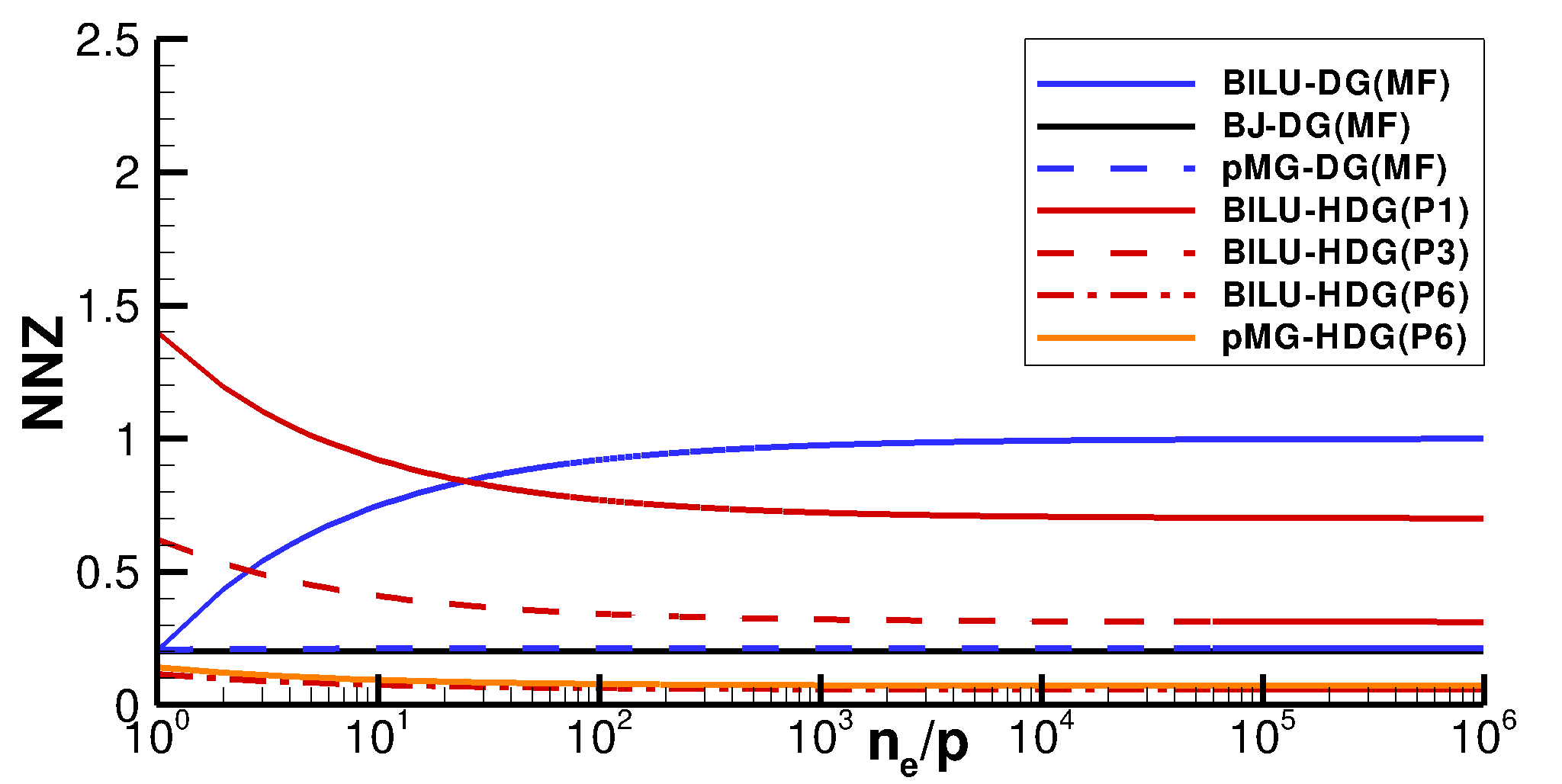}\label{fig:QuadLagrangeMem}}
\caption{Allocated number of non-zeros (NNZ) non-dimensionalized by the memory allocation of the DG Jacobian matrix. DG matrix-free solvers compared to HDG for different values of polynomial orders and preconditioning type. $p$-multigrid preconditioning ($p$MG) assumed to be that of Table~\ref{tab:MGSettings}.\label{fig:MemTot}}
\end{figure}
We now provide estimates of the memory footprint of the solver as well as the computational time spent evaluating the matrix operators. It is worth pointing out that, for DG, the matrix assembly time, as well as its operation count, is a function of the number of non-zeros of the matrix itself, while HDG involves the overhead costs of the static condensation and back solve. Considering a square, two-dimensional, bi-periodic domain made of quadrilateral elements, we obtain the results shown in Figure~\ref{fig:MemTot}, where the number of non-zeros (NNZ), non-dimensionalized by the number of nonzeros of the Jacobian arising from the DG discretization, is reported as a function of the number of elements per partition, $n_e/p$. It can be observed that:
\begin{enumerate}
    \item The memory footprint of DG, matrix-based as well as HDG solvers is always equal to that of a Jacobian matrix, and this value is a function of the polynomial order for HDG. This is due to the fact that the preconditioner is always evaluated using the same memory as the Jacobian matrix.
    \item For a matrix-free, DG discretization, the allocation involves only the preconditioner operator. When BJ is considered, NNZ reduces by 80\% with respect to the allocation of a full DG Jacobian. As $n_e/p\rightarrow 1$, the NNZ of the BILU solver approaches that of the element-wise block Jacobi, while for $n_e/p >> 1$ it tends to be that of a Jacobian matrix. This is due to the fact that as the domain is partitioned, the ILU(0) factorization is performed in the squared, partition-wise block of the iteration matrix and therefore, in a matrix-free fashion, the off-partition blocks can be neglected during the assembly phase.
    \item The $p$-multigrid ($p$MG) matrix-free preconditioning approach applied to a DG discretization, here assumed to be that of Table~\ref{tab:MGSettings}, requires a memory footprint in line with that of an element-wise block Jacobi method, as already observed in~\cite{franciolini2018p}. In fact, when using lower-order polynomial spaces with $k_{\ell}<<k$, the size of those matrices is considerably smaller than that of the finest space since they scale with $k^{2d}$.
    \item For HDG, only for high-order polynomials is NNZ reduced with respect to the iteration matrix of a DG method. For $k=6$, a memory footprint in line with that of a BJ, matrix-free approach is observed, while for $k=1,3$ the memory is considerably larger. It is worth pointing out that the use of full-order basis functions (Figure~\ref{fig:TriLagrangeMem}) and tensor-product basis functions (Figure~\ref{fig:QuadLagrangeMem}) only affects NNZ ratio in the HDG case: in particular, the memory footprint reduction when using tensor-product basis is larger due to the higher amount of element-wise unknowns compared to internal face unknowns.
    \item The NNZ ratio for HDG increases when $n_e/p\rightarrow 1$. The reason for this is that the face-to-elements ratio within the computational mesh of the domain partition increases.
\end{enumerate}
Finally, it is important to remark that for a matrix-free iterative solver employed in DG contexts, it is possible to optimize the matrix assembly evaluation to compute only the blocks required by the preconditioner. For example, for BJ, the evaluation of the off-diagonal blocks of the Jacobian can be neglected, while for $p$MG matrix-free with BJ on the finest space, the off-diagonal blocks could be computed at a reduced polynomial order consistent with that of the coarser spaces.


\section{Numerical results on a model test case} \label{sec:results}

We present numerical experiments to assess the performance of the HDG discretizations in comparison to DG. First, NS solutions of a vortex transported by uniform flow at $M=0.05$ and $Re=100$ are reported. The objective is i) to show the convergence rates of the solver both in space and time; ii) to investigate the effects of grid refinement for the approximate-inherited approach proposed for HDG, providing mesh-independent convergence rates; and iii) compare the effects of polynomial order, time step size and space discretization on the parallel performance of the solution strategy.

\subsection{Test case description}

The test case is a modified version of the VI1 case studied in the 5\textsuperscript{th} International Workshop on High Order CFD Methods~\cite{5HHW}, and consists of a two-dimensional mesh on the domain $(x,y) \in [0,0.1]{\times}[0,0.1]$ with periodic boundary conditions on each side. The flow initialisation involves the definition of the following state
\begin{equation}
\begin{aligned}
u&=U_{\infty}\left(1-\beta \left(\dfrac{y-Y_c}{R}\right)e^{-r^2/2}\right)\\[2mm]
v&=U_{\infty}\beta \left(\dfrac{x-X_c}{R}\right)e^{-r^2/2}\\[2mm]
T&=T_{\infty}- \left(\dfrac{U_{\infty}^2\beta^2}{2C_p}\right)e^{-r^2}
\end{aligned}
\end{equation}
with the heat capacity at constant pressure $C_p=R_{\textrm{gas}}\gamma/(\gamma-1)$, the non dimensional distance to the initial vortex core position $r=\sqrt{(x-X_c)^2+(y-Y_c)^2}/R$, being $X_c,Y_c$ the coordinates of the vortex center, and the free stream velocity $U_{\infty}=M_{\infty}\sqrt{\gamma R_{\textrm{gas}}T_{\infty}}$. The fluid pressure $p$, the temperature $T$ and density $\rho$ are prescribed to ensure a steady solution of the problem without uniform flow transport, \ie $\rho_{\infty}=p_{\infty}/R_{\textrm{gas}}T_{\infty}$, $\rho=\rho_{\infty}(T/T_{\infty})^{1/(\gamma-1)}$, $p=\rho R_{\textrm{gas}} T$. The parameters where chosen such that $M_{\infty}=0.05$, $\beta=1/50$ and $R=0.005$. In contrast to the inviscid-flow case studied in the workshop, the governing equations in the present study are Navier-Stokes, with $Re=100$ based on the domain size.

\subsection{Assessment of the solution accuracy} 
\begin{figure}[t!]
\centering
\includegraphics[width=0.38\textwidth]{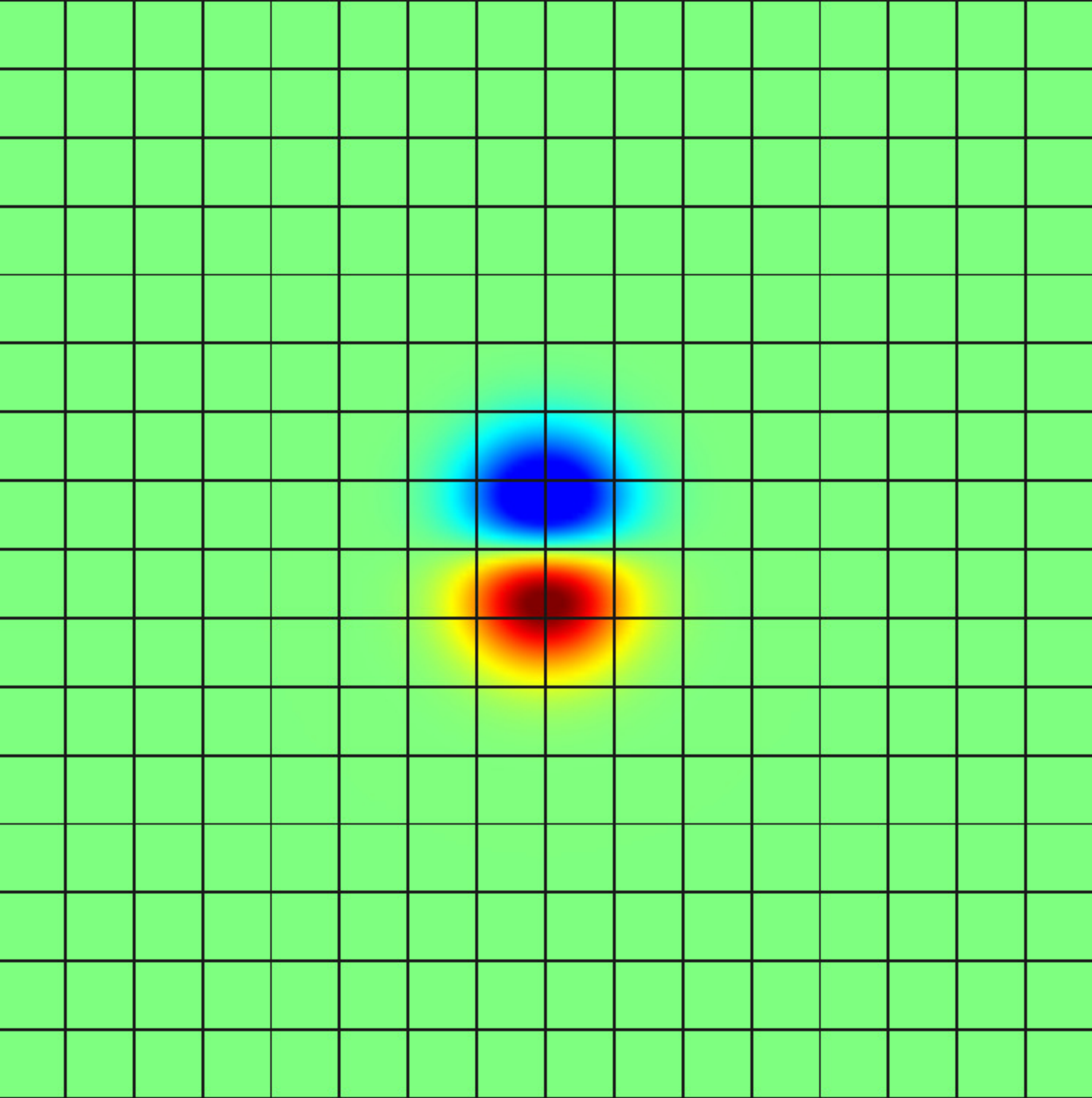}\qquad
\includegraphics[width=0.38\textwidth]{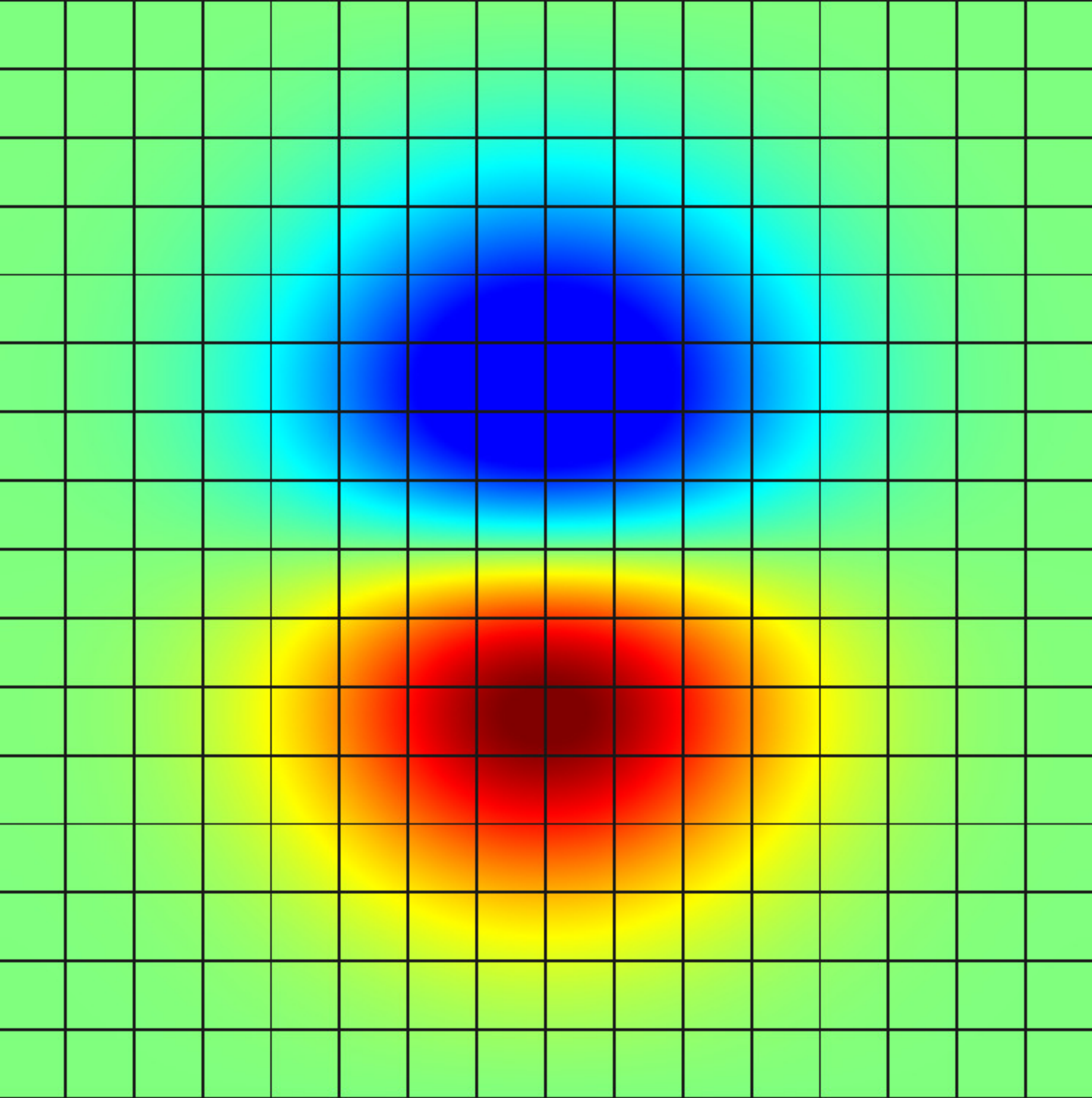}
\caption{Convected vortex at $Re=100$, $M=0.05$. Mach number contours. Solution at $t=0$ (left) and $t=T$ (right).}
\label{fig:Vortex}
\end{figure}

Numerical experiments have been performed to assess the output error, both in space and time. The meshes here employed were obtained using regular quadrilaterals. The mesh density ranges from $2{\times}2$ to $64{\times}64$, while the polynomial order range $k\in \{1,2,3,4,5,6\}$. The $L_2$ state error was computed relative to the solution on a $128{\times}128$, $\mathbb{P}_6$ space discretization, after one convective period $T$. The contour plot of the solutions at the initial and final states are shown in Figure~\ref{fig:Vortex}.
\begin{table}[htbp!]
\centering
\begin{tabular}{|c|c||c|c||c|c||c|c|}
\hline
 &  & \multicolumn{2}{c||}{DG} & \multicolumn{2}{c||}{$m$HDG} & \multicolumn{2}{c|}{$p$HDG} \\\hline
order & grid & $\|err\|_{L_2}$ & $k$ & $\|err\|_{L_2}$ & $k$ & $\|err\|_{L_2}$ & $k$ \\\hline\hline
\multirow{4}{*}{$\mathbb{P}_1$} & 8 & 4.0938E-07 &  & 4.20E-07 &  & 4.1798E-07 & \\
 & 16 & 1.4796E-07 & 1.468 & 1.4514E-07 & 1.533 & 1.4406E-07 & 1.537 \\
 & 32 & 3.0795E-08 & 2.264 & 2.8355E-08 & 2.356 & 2.8020E-08 & 2.362 \\
 & 64 & 5.5247E-09 & 2.479 & 4.7822E-09 & 2.568 & 4.6898E-09 & 2.579 \\\hline\hline
\multirow{6}{*}{$\mathbb{P}_2$} & 2 & 5.9266E-07 &  & 6.22E-07 &  & 6.2124E-07 & \\
 & 4 & 2.7443E-07 & 1.111 & 2.7344E-07 & 1.186 & 2.7306E-07 & 1.186 \\
 & 8 & 3.2505E-08 & 3.078 & 3.2777E-08 & 3.060 & 3.2344E-08 & 3.078 \\
 & 16 & 2.2983E-09 & 3.822 & 2.7457E-09 & 3.577 & 2.6311E-09 & 3.620 \\
 & 32 & 2.3510E-10 & 3.289 & 3.4213E-10 & 3.005 & 3.1517E-10 & 3.061 \\
 & 64 & 3.4853E-11 & 2.754 & 6.3410E-11 & 2.432 & 5.5951E-11 & 2.494 \\\hline\hline
\multirow{6}{*}{$\mathbb{P}_3$} & 2 & 3.3987E-07 &  & 3.31E-07 &  & 3.3149E-07 & \\
 & 4 & 3.2850E-08 & 3.371 & 3.4696E-08 & 3.255 & 3.4275E-08 & 3.274 \\
 & 8 & 1.5714E-09 & 4.386 & 1.7405E-09 & 4.317 & 1.7008E-09 & 4.333 \\
 & 16 & 8.1804E-11 & 4.264 & 7.9918E-11 & 4.445 & 7.8935E-11 & 4.429 \\
 & 32 & 7.2397E-12 & 3.498 & 7.2689E-12 & 3.459 & 7.2173E-12 & 3.451 \\\hline\hline
\multirow{5}{*}{$\mathbb{P}_4$} & 2 & 1.1210E-07 &  & 1.18E-07 &  & 1.1741E-07 & \\
 & 4 & 5.2542E-09 & 4.415 & 5.4176E-09 & 4.446 & 5.3668E-09 & 4.451 \\
 & 8 & 1.1772E-10 & 5.480 & 1.9915E-10 & 4.766 & 1.9838E-10 & 4.758 \\
 & 16 & 6.5563E-12 & 4.166 & 7.3117E-12 & 4.768 & 7.1101E-12 & 4.802 \\\hline\hline
\multirow{5}{*}{$\mathbb{P}_5$} & 2 & 4.2677E-08 &  & 4.50E-08 &  & 4.4730E-08 & \\
 & 4 & 6.6523E-10 & 6.003 & 7.8204E-10 & 5.848 & 7.7398E-10 & 5.853 \\
 & 8 & 1.1519E-11 & 5.852 & 9.6028E-11 & 3.026 & 9.6035E-11 & 3.011 \\
 & 16 & 5.0750E-12 & 1.183 & 4.8543E-12 & 4.306 & 4.8326E-12 & 4.313 \\\hline\hline
\multirow{5}{*}{$\mathbb{P}_6$} & 2 & 1.4170E-08 &  & 1.35E-08 &  & 1.3510E-08 & \\
 & 4 & 9.7090E-11 & 7.189 & 3.4373E-10 & 5.298 & 3.4301E-10 & 5.300 \\
 & 8 & 5.1249E-12 & 4.244 & 2.4622E-11 & 3.803 & 2.4821E-11 & 3.789 \\
 & 16 & 5.0800E-12 & 0.013 & 5.1008E-12 & 2.271 & 5.0943E-12 & 2.285 \\\hline
\end{tabular}
\caption{$L_2$ solution error. Laminar vortex test case at $Re=100$, $M=0.05$. Convergence rates for the DG and HDG discretizations.}
\label{tab:VOR_CONVP}
\end{table}
Table~\ref{tab:VOR_CONVP} reports space discretization errors. The tests were performed using a very small time step size, $T/\Delta t=4000$, and the ESDIRK3 scheme to ensure a negligible time discretization error, with an absolute tolerance on the non-linear system of $10^{-10}$, and a relative tolerance of $10^{-5}$ on GMRES. Even though DG suffers less than HDG from pre-asympthotic behaviour on such a smooth solution, all three implementations show comparable error levels and converge with the theoretical convergence rates for every polynomial approximation shown. As a consequence of such analysis, and considering that both the DG and HDG implementations share the same code base, we will consider only the CPU time as a measure of the time-to-solution efficiency.

Regarding the time integration scheme, the convergence rates of ESDIRK3 are also reported in Table~\ref{tab:VOR_CONVT}, and these were obtained using the $16{\times}16$ grid, with $\mathbb{P}_6$ polynomials for the three space discretization strategies.
\begin{table}[htbp!]
\centering
\begin{tabular}{|c||c|c||c|c||c|c|}
\hline
 & \multicolumn{2}{c||}{DG} &  \multicolumn{2}{c||}{$m$HDG}&  \multicolumn{2}{c|}{$p$HDG} \\\hline
$T/\Delta t$ & $\|err\|_{L_2}$ & $k$ & $\|err\|_{L_2}$ & $k$ & $\|err\|_{L_2}$ & $k$ \\\hline\hline
4 & 5.3895E-07 &  & 5.3898E-07 &  & 5.3898E-07 & \\
10 & 1.3656E-07 & 1.498 & 1.3657E-07 & 1.498 & 1.3657E-07 & 1.498 \\
20 & 2.5830E-08 & 2.402 & 2.5788E-08 & 2.405 & 2.5788E-08 & 2.405 \\
40 & 3.7729E-09 & 2.775 & 3.7684E-09 & 2.775 & 3.7684E-09 & 2.775 \\
100 & 2.6181E-10 & 2.912 & 2.6049E-10 & 2.916 & 2.6050E-10 & 2.916 \\\hline
\end{tabular}
\caption{Laminar vortex test case at $Re=100$, $M=0.05$. Temporal convergence rates for the DG, $m$HDG and $p$HDG discretizations, using the ESDIRK3 scheme.}
\label{tab:VOR_CONVT}
\end{table}
The theoretical third-order convergence rate of the ESDIRK3 time integration scheme can be observed for all three space discretizations with comparable error levels.

\subsection{Assessment of the approximate-inherited multigrid approach for HDG}
\begin{figure}[t!]
\centering
\subfigure[]{\includegraphics[width=0.22\textwidth]{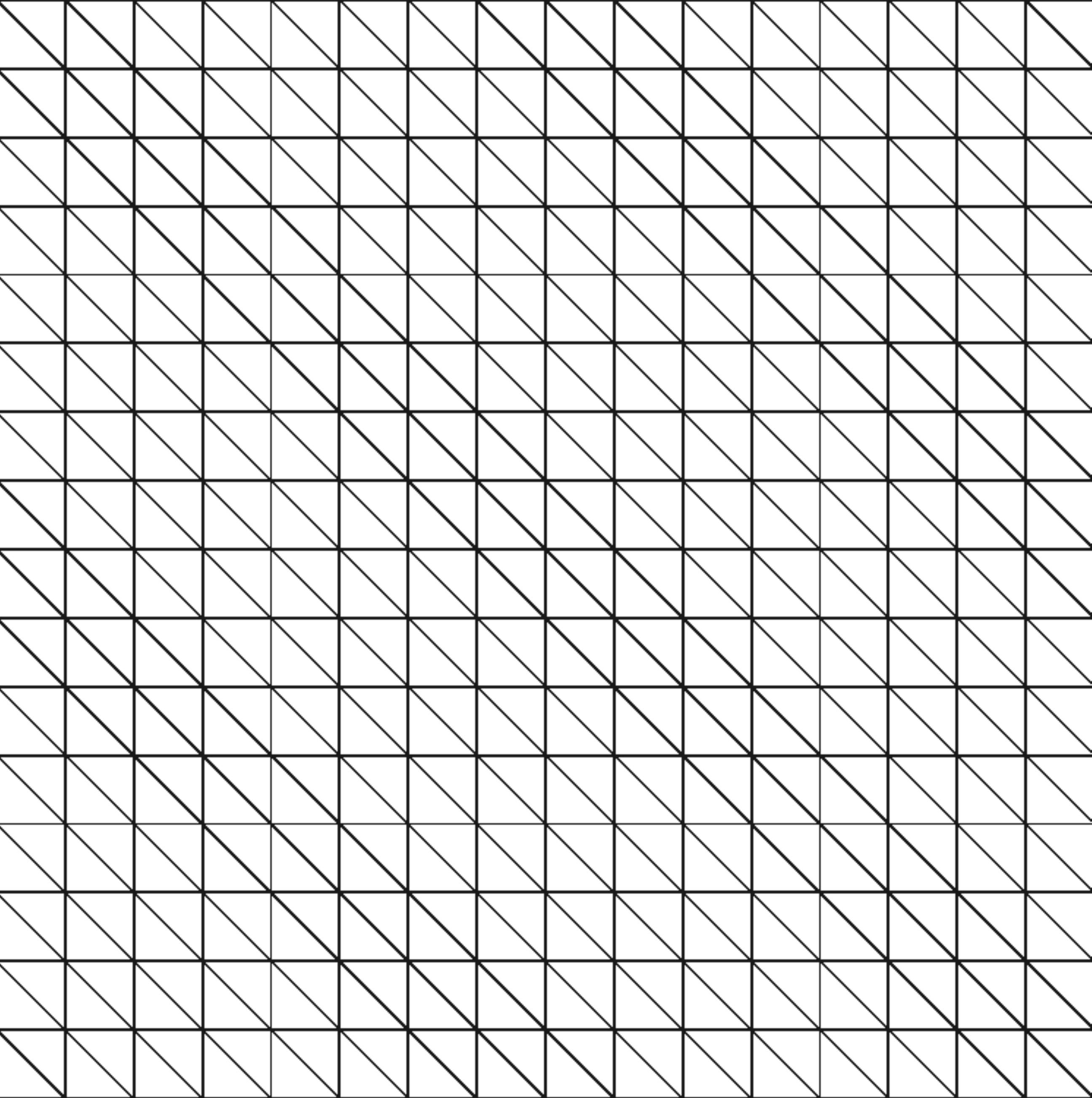}}\quad
\subfigure[]{\includegraphics[width=0.22\textwidth]{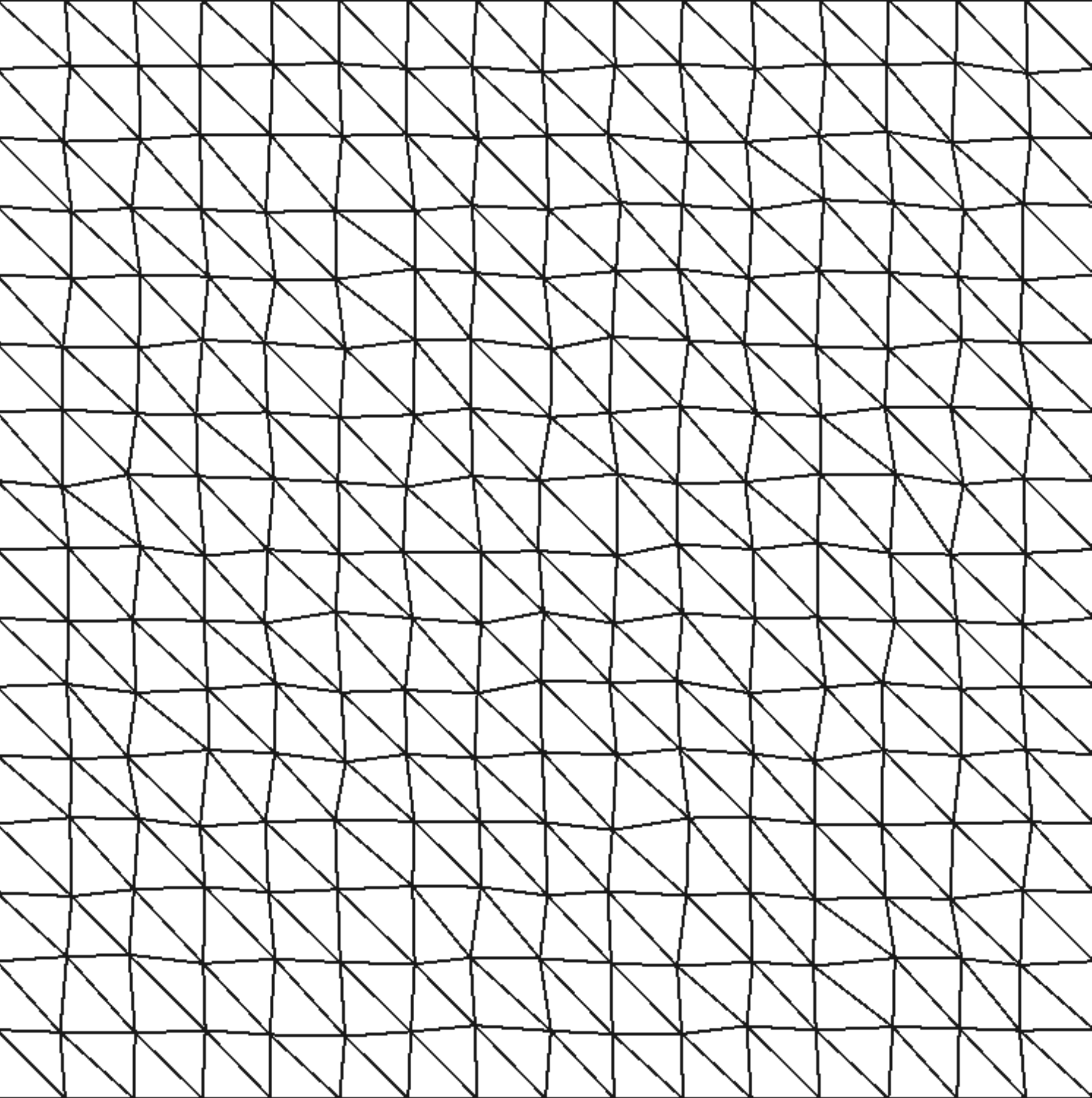}}\quad
\subfigure[]{\includegraphics[width=0.22\textwidth]{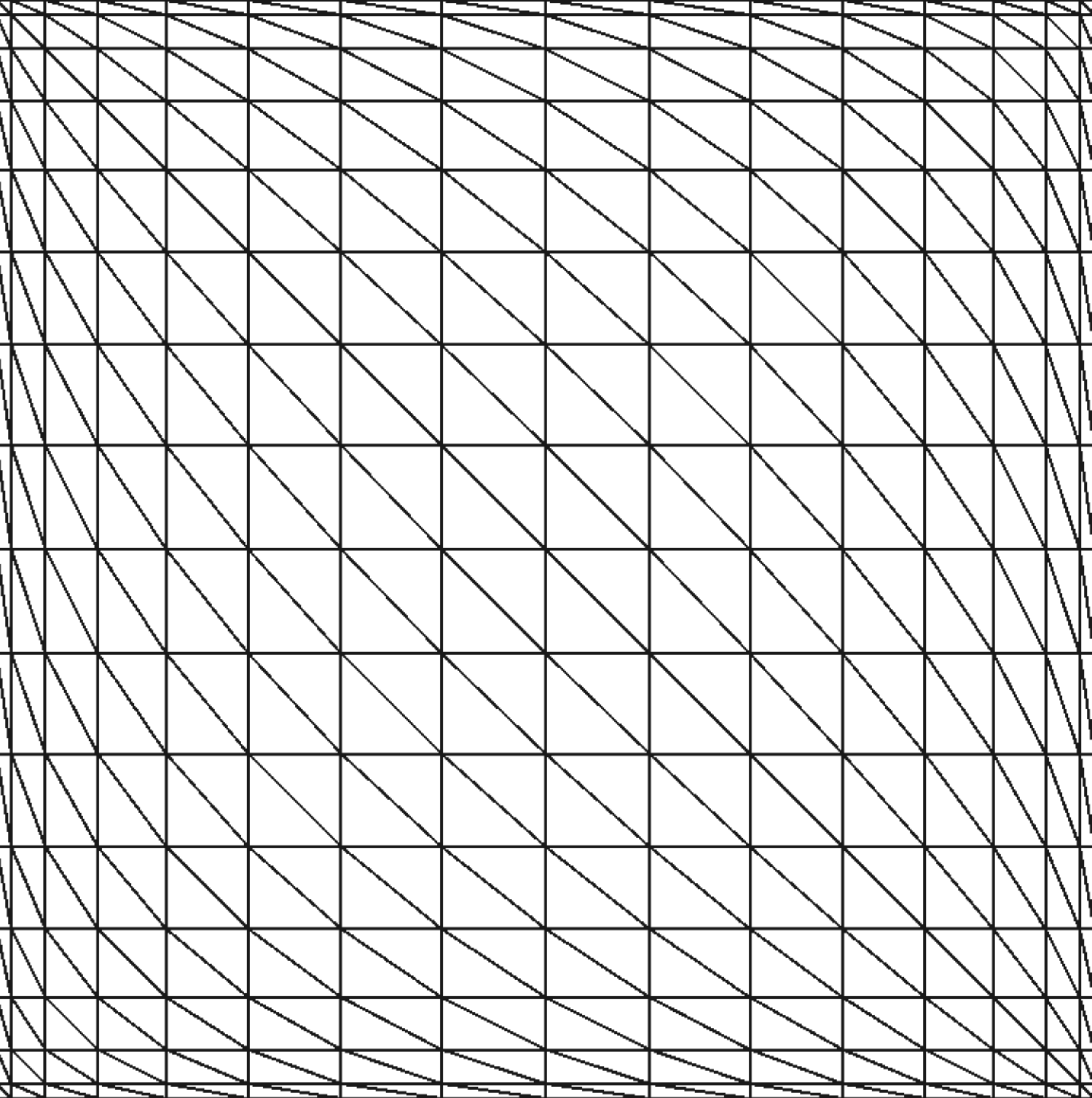}}\quad
\subfigure[]{\includegraphics[width=0.22\textwidth]{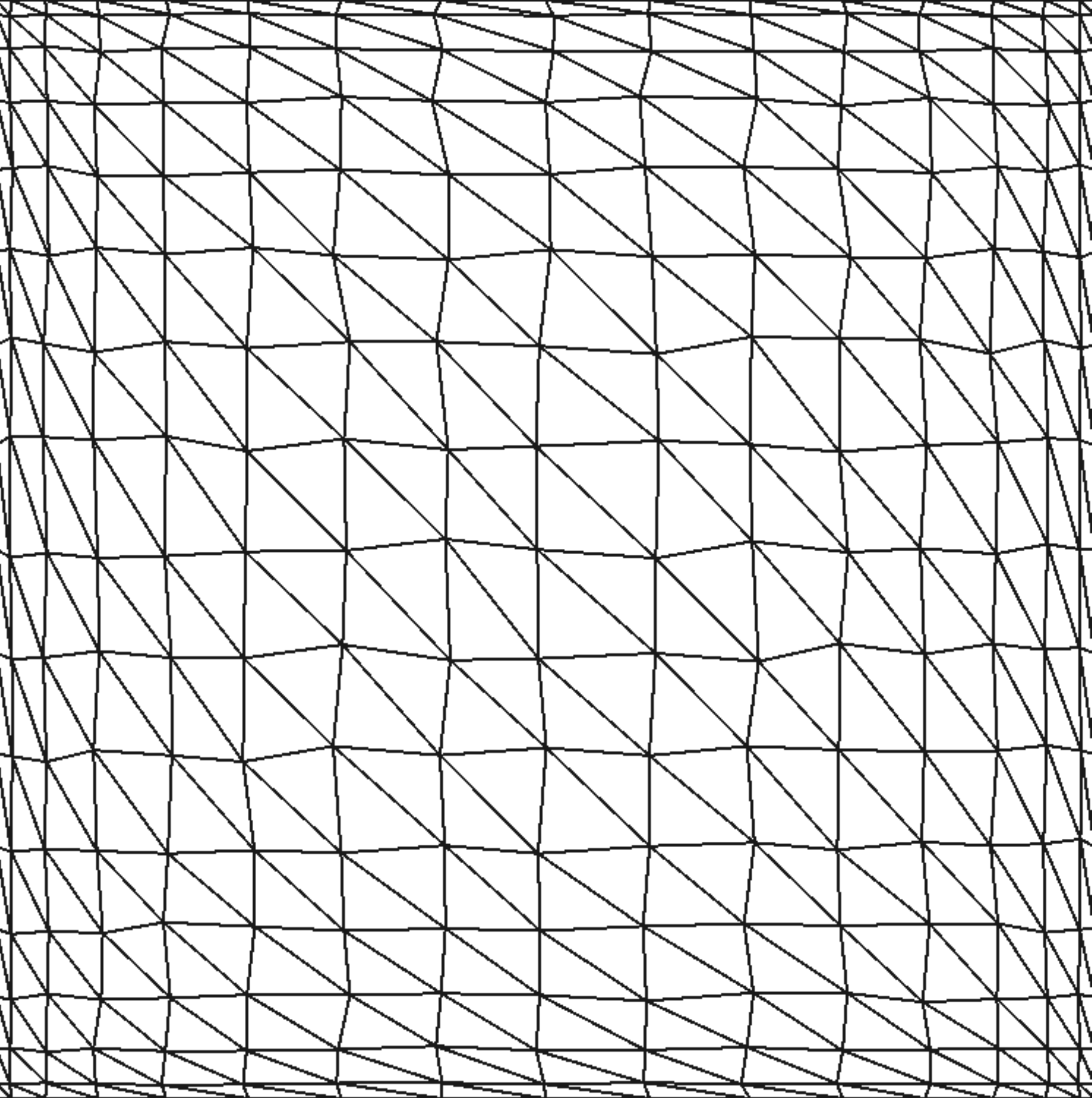}}\quad
\subfigure[]{\includegraphics[width=0.22\textwidth]{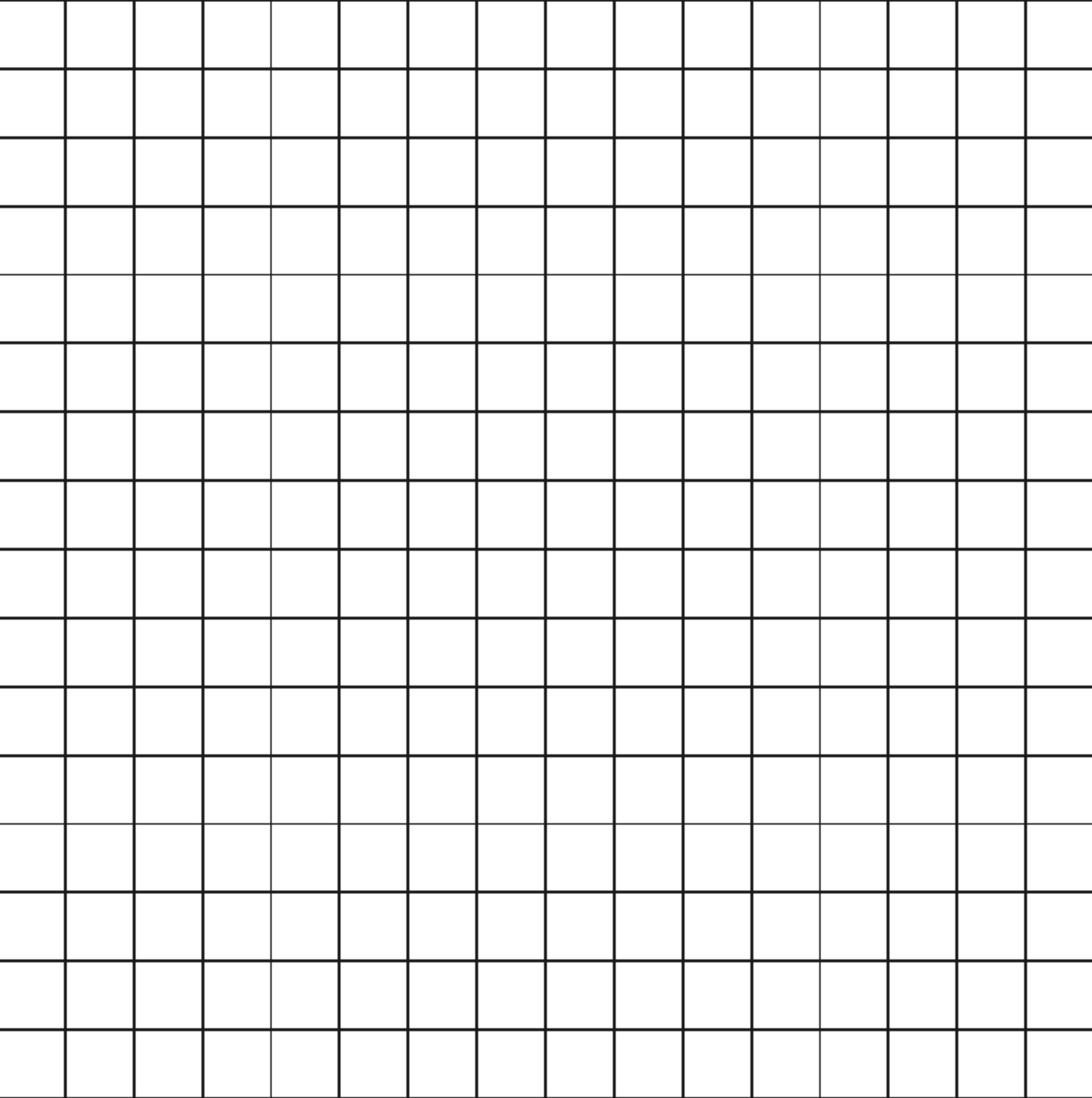}}\quad
\subfigure[]{\includegraphics[width=0.22\textwidth]{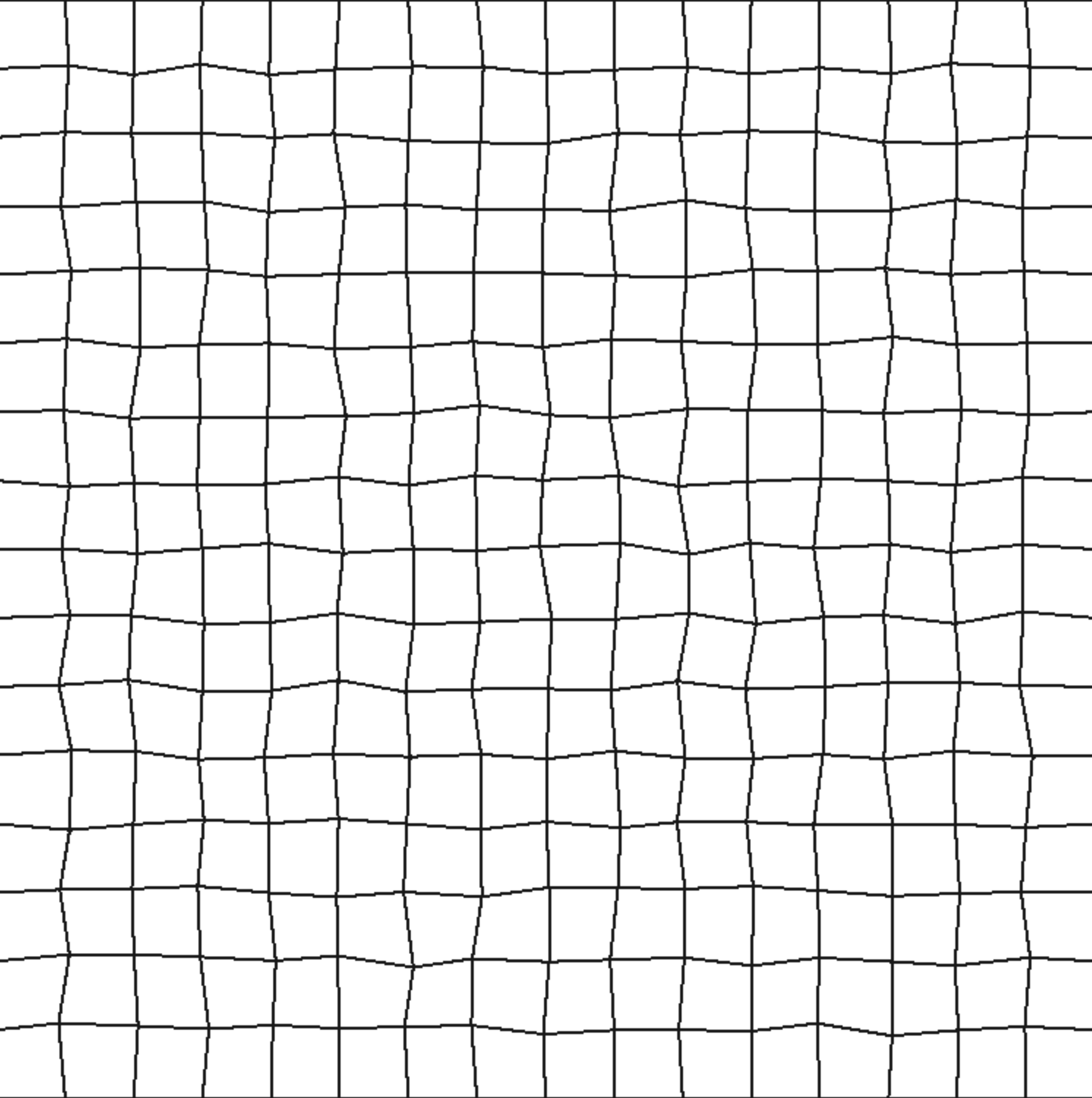}}\quad
\subfigure[]{\includegraphics[width=0.22\textwidth]{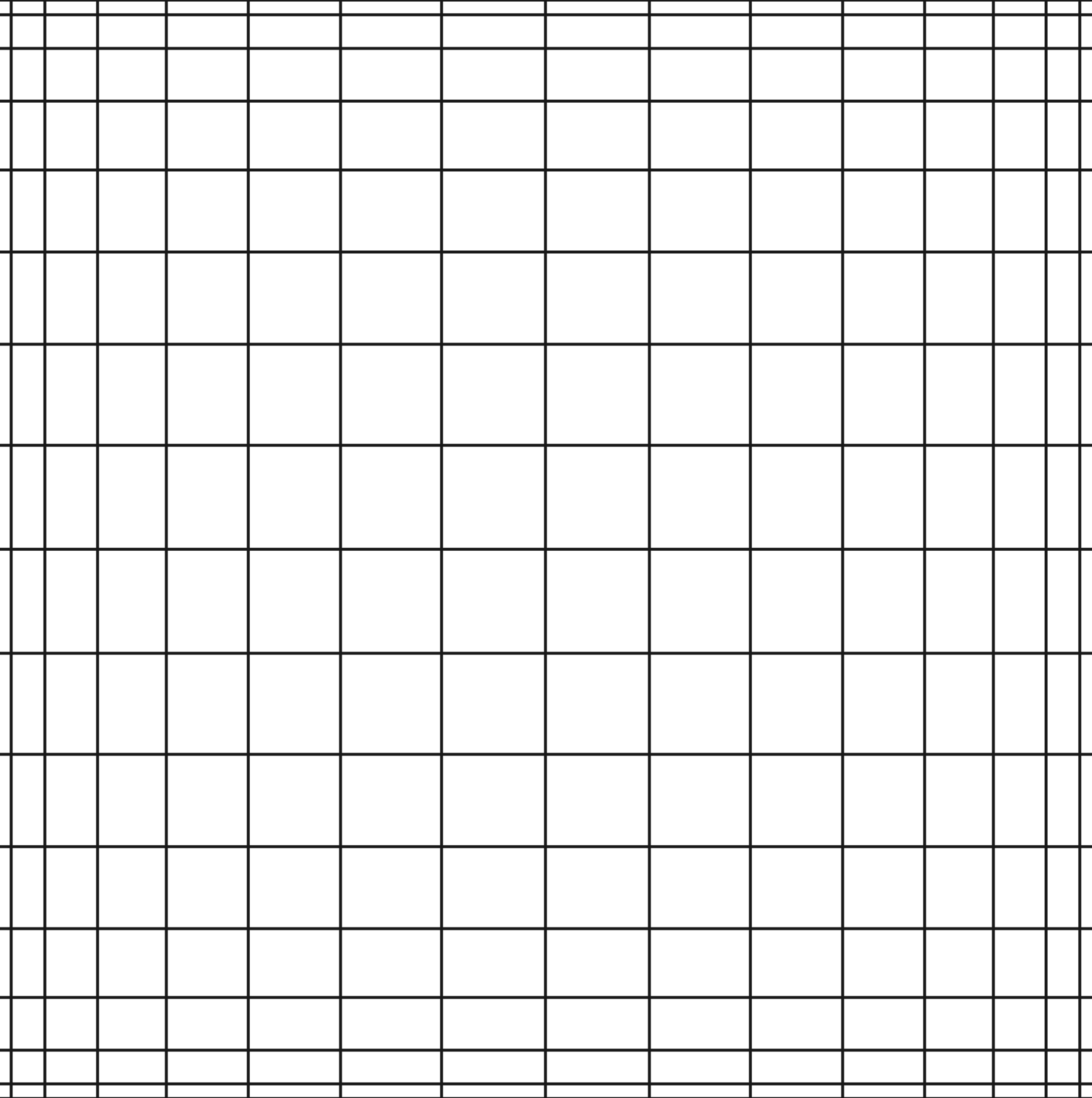}}\quad
\subfigure[]{\includegraphics[width=0.22\textwidth]{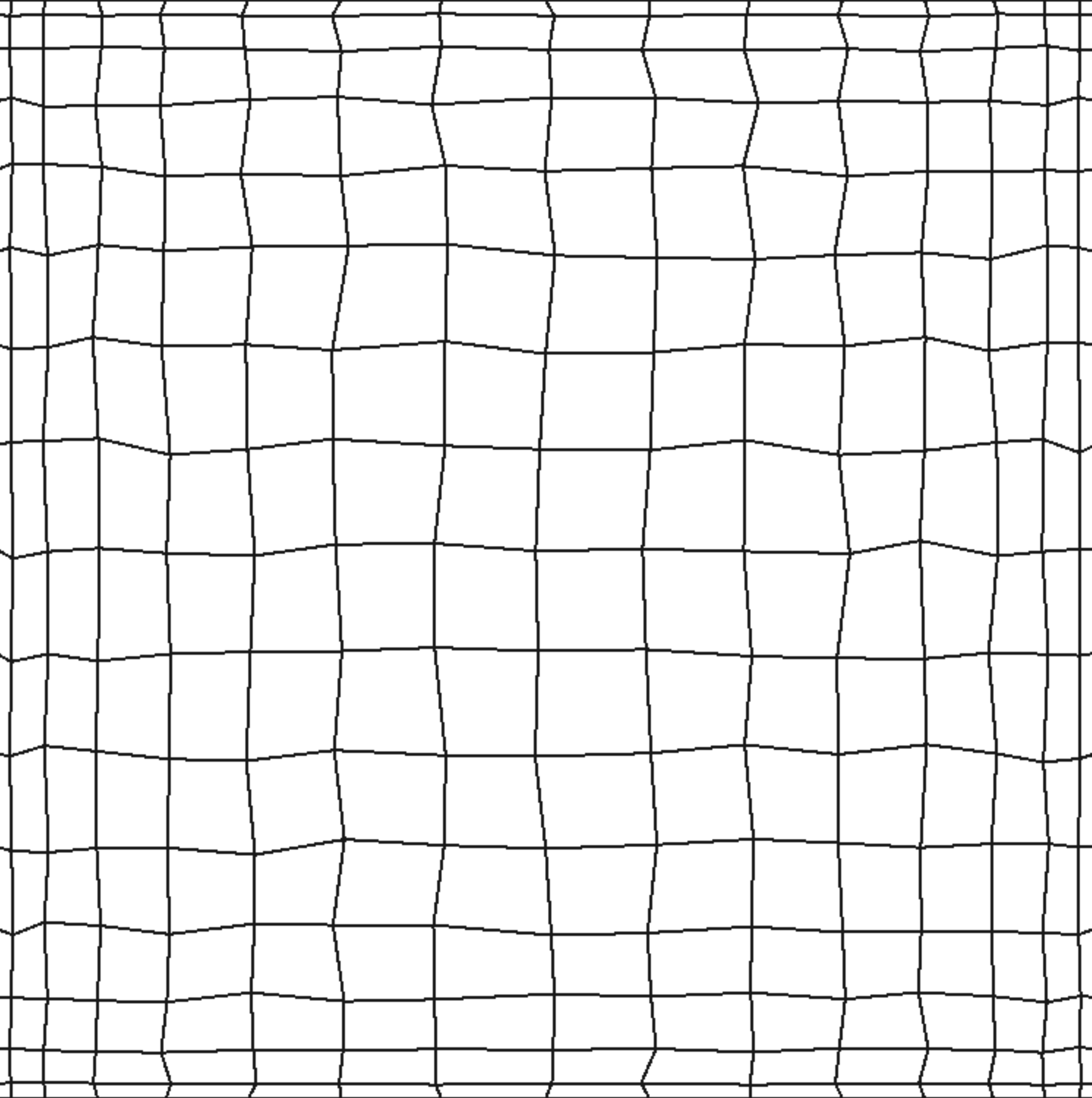}}
\caption{Convected vortex at $Re=100$, $M=0.05$. Example of computatinal meshes involving, for the tri- and quad-, i) regular elements; ii) randomly distorted elements; iii) regular and clustered elements; and iv) clustered-distorted elements.}
\label{fig:HDGGrids}
\end{figure}
To demonstrate the efficacy of the multigrid approach proposed herein for HDG, we report numerical experiments obtained by reducing the mesh element size for different element types. Four mesh sequences are considered in the study, obtained as i) regular elements; ii) randomly distorted elements; iii) regular and clustered elements; and iv) clustered-distorted elements. The study was performed on both triangular and quadrilateral elements, see Figure~\ref{fig:HDGGrids}. We remark that the random mesh perturbation was limited to the 10\% of the minimum dimension of the element, and that the clustering has been obtained by placing the mesh element nodes using Gauss-Lobatto rules. A full multigrid strategy has been employed for the study. The strategy combines three multigrid levels, and for each of them a BILU-preconditioned GMRES smoother is considered. To avoid the strategy being influenced by the domain decomposition, all the computations are here performed in serial. Three smoothing iterations have been performed in each level, while 400 iterations were employed on the coarsest level to ensure that the results are not polluted by a lack of coarse level resolution. This configuration was found to optimize the solver to deal with serial computations and it is coherent to what has been previously reported in the literature, see~\cite{franciolini2018p}.

\begin{table}[t!]
\centering
\begin{tabular}{|c|cc|cc|cc|cc|}
\hline
 & \multicolumn{2}{c|}{Reg Tri} & \multicolumn{2}{c|}{Dis Tri} & \multicolumn{2}{c|}{Reg Grad Tri} & \multicolumn{2}{c|}{Dis Grad Tri} \\\hline
$n_e$ & $IT_a$ & $\rho_a$ & $IT_a$ & $\rho_a$ & $IT_a$ & $\rho_a$ & $IT_a$ & $\rho_a$ \\ \hline
16$^2 \cdot 2$ & 3.000 & 0.0060 & 2.833 & 0.0056 & 3.000 & 0.0120 & 3.000 & 0.0131 \\
32$^2 \cdot 2$ & 3.000 & 0.0123 & 3.000 & 0.0119 & 3.500 & 0.0245 & 3.333 & 0.0195 \\
64$^2 \cdot 2$ & 2.500 & 0.0058 & 3.000 & 0.0132 & 3.667 & 0.0315 & 4.000 & 0.0422 \\
128$^2 \cdot 2$ & 2.333 & 0.0047 & 2.667 & 0.0086 & 3.500 & 0.0290 & 4.167 & 0.0560 \\\hline\hline
 & \multicolumn{2}{c|}{Reg Quad} & \multicolumn{2}{c|}{Dis Quad} & \multicolumn{2}{c|}{Reg Grad Quad} & \multicolumn{2}{c|}{Dis Grad Quad} \\\hline
$n_e$ & $IT_a$ & $\rho_a$ & $IT_a$ & $\rho_a$ & $IT_a$ & $\rho_a$ & $IT_a$ & $\rho_a$ \\ \hline
16$^2$ & 2.833 & 0.0053 & 2.333 & 0.0033 & 3.000 & 0.0069 & 3.000 & 0.0057 \\
32$^2$ & 2.833 & 0.0101 & 2.833 & 0.0081 & 3.000 & 0.0087 & 3.000 & 0.0096 \\
64$^2$ & 3.167 & 0.0183 & 3.000 & 0.0153 & 3.833 & 0.0386 & 3.667 & 0.0336 \\
128$^2$ & 3.333 & 0.0246 & 3.500 & 0.0273 & 4.333 & 0.0637 & 4.500 & 0.0659 \\\hline
\end{tabular}
\caption{Laminar vortex test case at $Re=100$, $M=0.05$. $h$-independence test on tri-element (top) and quad-element (bottom) meshes. of a FGMRES(MG$_{\textrm{full}}$) solver built on three levels. GMRES(BILU) smoothers with 400 iterations on the coarsest level $\ell=2$. 3 smoothing iterations were employed for $\ell=\{0,1\}$. }
\label{tab:h-ind-tri-quad}
\end{table}
Table~\ref{tab:h-ind-tri-quad} report the results on triangular and quadrilateral mesh elements. The numerical experiment consists of one time period such that $T/\Delta t=10$ using the ESDIRK3 scheme, with $T$ the convective period of the problem. The average number of iterations as well as the average convergence rate (CR) $\rho$ are reported. The CR is defined as $\rho = (r_{IT}/r_0)^{1/IT}$
with $IT$ the average number of GMRES iterations, $r_0$ and $r_{IT}$ the residuals at the first and $IT$\textsuperscript{th} iteration respectively. In all of the numerical experiments, the $p$-multigrid strategy appears to work optimally as the number of iterations slightly grows by increasing the number of mesh elements, even for distorted and graded mesh sequences.


\subsection{Evaluation of the solver efficiency}


We report results of numerical experiments devoted to assess the performance of the $p$-multigrid preconditioning strategy for HDG in comparison to other operators, as well as state-of-the-art preconditioned DG discretizations. In particular, we take as a reference the matrix-based, BILU preconditioned DG discretization and we aim at reporting insights on the computational time of the solution, in order to provide an overall idea of the computational efficiency of the solver. The space discretization relies on the $16{\times}16$ mesh made by regular quadrilaterals and two polynomial orders, \emph{i.e.} $k=\{3;6\}$. As for the time discretization, non dimensional time steps of $\Delta t = \{1;0.1\}$ are employed. In both cases, a single time step is computed, which corresponds to three non-linear problems. We observe that our Newton solver converges in two iterations, which means that the CPU time and the average number of iterations is evaluated considering a total of six linear system solutions. Each linear system is solved up to a non-preconditioned relative linear tolerance of $10^{-5}$, while an absolute tolerance of $10^{-10}$ was used for the nonlinear solver.

Special attention is hereby given to the parallel efficiency of the computations, both in terms of CPU time and average number of GMRES iterations. To this extent, the ideas reported in~\cite{franciolini2018p} on the choice of the number of levels and the smoothing type have been proposed. In fact, despite the fact that that work focused on incompressible flow problems, a very similar behaviour of the solver has been presently observed and thus we explicitly refer to that work for a more in-depth discussion about the effects of the smoothing. We here report only the general idea: a scalable multigrid strategy to precondition systems arising from the Navier--Stokes equations can be obtained by the use of simple BJ-preconditioned GMRES smoothers for all of the levels except the coarse one, where more powerful preconditioners can be cheaply introduced on the smoothers due to the low computational cost of the coarse matrix factorization. 
\begin{table}[t!]
\footnotesize
\centering
\begin{tabular}{|c|c|c|c|c|}
\hline
Level & Order & Solver & Preconditioner & Iterations \\\hline\hline
1 & 6 & GMRES & BJ & 10 \\\hline
2 & 2 & GMRES & BJ & 10 \\\hline
3 & 1 & GMRES & BILU & 30 \\\hline
\end{tabular}
\caption{Computational settings for the $p$-multigrid preconditioning strategy employed within the paper.}
\label{tab:MGSettings}
\end{table}

In the numerical experiments we rely on a three-level multigrid strategy. In both the cases, $k_{1,2}=\{2,1\}$ have been used on the coarser spaces smoothed with BJ- and BILU-preconditioned GMRES solvers, respectively. $\{10,30\}$ smoothing iterations have been employed, which were found to be sufficient to provide optimal results both in serial and in parallel computations for HDG and DG as well. On the finest space, 10 smoothing iterations of GMRES(BJ) were used. The computational settings are summarized in Table~\ref{tab:MGSettings}. We remark that those settings have been found to be enough computationally efficient for all the numerical experiments reported throughout the paper.

\begin{table}[t!]
\footnotesize
\centering
\begin{tabular}{|c||ccc|ccc||ccc|ccc|}
\hline
Discr. & \multicolumn{6}{c||}{$k=3$, $\Delta t=1$} & \multicolumn{6}{c|}{$k=3$, $\Delta t=0.1$}\\\hline
Solver & \multicolumn{3}{c|}{BILU-DG} & \multicolumn{3}{c||}{MG$_{\textrm{full}}$-DG} & \multicolumn{3}{c|}{BILU-DG} & \multicolumn{3}{c|}{MG$_{\textrm{full}}$-DG}\\\hline
$n_p$ & Time & $E$ & $IT_a$ & $SU_{MB}$ & $E$ & $IT_a$ & Time & $E$ & $IT_a$ & $SU_{MB}$ & $E$ & $IT_a$ \\
1 & 38.52 &  & 81.00 & 1.03 &  & 3.50 & 23.51 &  & 27.83 & 0.88 &  & 2.17 \\
2 & 27.53 & 0.70 & 137.33 & 1.43 & 0.97 & 3.50 & 15.12 & 0.78 & 53.17 & 1.08 & 0.96 & 2.17 \\
4 & 15.10 & 0.64 & 154.83 & 1.51 & 0.94 & 3.50 & 7.68 & 0.77 & 55.33 & 1.06 & 0.92 & 2.17 \\
8 & 8.72 & 0.55 & 181.17 & 1.58 & 0.85 & 3.67 & 4.15 & 0.71 & 65.50 & 1.04 & 0.84 & 2.17 \\
16 & 6.03 & 0.40 & 229.50 & 1.78 & 0.69 & 4.00 & 2.23 & 0.66 & 72.33 & 0.99 & 0.74 & 2.17 \\
32 & 5.82 & 0.21 & 284.50 & 2.09 & 0.42 & 4.67 & 1.57 & 0.47 & 81.83 & 0.96 & 0.51 & 2.17 \\\hline\hline
Solver & \multicolumn{3}{c|}{BILU-$m$HDG} & \multicolumn{3}{c||}{MG$_{\textrm{full}}$-$m$HDG} & \multicolumn{3}{c|}{BILU-$m$HDG} & \multicolumn{3}{c|}{MG$_{\textrm{full}}$-$m$HDG}\\\hline
$n_p$ & $SU_{MB}$ & $E$ & $IT_a$ & $SU_{MB}$ & $E$ & $IT_a$ & $SU_{MB}$ & $E$ & $IT_a$ & $SU_{MB}$ & $E$ & $IT_a$ \\
1 & 1.49 &  & 45.17 & 1.23 &  & 2.50 & 0.96 &  & 20.33 & 0.79 &  & 2.00 \\
2 & 1.75 & 0.82 & 77.17 & 1.54 & 0.88 & 2.50 & 1.03 & 0.83 & 36.00 & 0.88 & 0.87 & 2.00 \\
4 & 1.63 & 0.70 & 95.50 & 1.50 & 0.78 & 2.50 & 0.91 & 0.72 & 41.00 & 0.80 & 0.78 & 2.00 \\
8 & 1.67 & 0.62 & 113.17 & 1.56 & 0.70 & 2.50 & 0.90 & 0.66 & 46.83 & 0.78 & 0.70 & 2.00 \\
16 & 1.76 & 0.47 & 138.67 & 1.75 & 0.57 & 2.83 & 0.79 & 0.54 & 52.67 & 0.68 & 0.56 & 2.00 \\
32 & 1.93 & 0.27 & 183.50 & 2.21 & 0.37 & 3.50 & 0.81 & 0.39 & 64.50 & 0.70 & 0.42 & 2.00 \\\hline\hline
Solver & \multicolumn{3}{c|}{BILU-$p$HDG} & \multicolumn{3}{c||}{MG$_{\textrm{full}}$-$p$HDG} & \multicolumn{3}{c|}{BILU-$p$HDG} & \multicolumn{3}{c|}{MG$_{\textrm{full}}$-$p$HDG}\\\hline
$n_p$ & $SU_{MB}$ & $E$ & $IT_a$ & $SU_{MB}$ & $E$ & $IT_a$ & $SU_{MB}$ & $E$ & $IT_a$ & $SU_{MB}$ & $E$ & $IT_a$  \\
1 & 2.45 &  & 44.33 & 1.95 &  & 2.00 & 1.62 &  & 44.33 & 1.19 &  & 2.00 \\
2 & 2.83 & 0.81 & 75.17 & 2.48 & 0.89 & 2.00 & 1.75 & 0.84 & 75.17 & 1.36 & 0.89 & 2.00 \\
4 & 2.60 & 0.68 & 94.17 & 2.26 & 0.74 & 2.50 & 1.55 & 0.73 & 94.17 & 1.22 & 0.79 & 2.00 \\
8 & 2.57 & 0.58 & 112.33 & 2.34 & 0.67 & 2.50 & 1.51 & 0.66 & 112.33 & 1.19 & 0.71 & 2.00 \\
16 & 2.67 & 0.44 & 136.83 & 2.60 & 0.53 & 2.67 & 1.29 & 0.52 & 136.83 & 1.05 & 0.58 & 2.00 \\
32 & 2.59 & 0.22 & 182.83 & 3.05 & 0.32 & 3.50 & 1.38 & 0.40 & 182.83 & 1.05 & 0.41 & 2.00 \\\hline
\end{tabular}
\caption{Computational efficiency comparison using DG and HDG discretizations. Laminar vortex test case at $Re=100$, $M=0.05$, discretized using $16{\times}16$ mesh with $\mathbb{P}_3$ polynomials. Two time steps, $\Delta t=\{1;0.1\}$ using the ESDIRK3 scheme are reported. $SU_{MB}$ stands for the speed-up factor relative to the DG, matrix-based, BILU-preconditioned computation, $E$ is the parallel efficiency and $IT_a$ the average number of GMRES iterations.}
\label{tab:P3ViscVortPar}
\end{table}
\begin{table}[t!]
\footnotesize
\centering
\begin{tabular}{|c||ccc|ccc||ccc|ccc|}
\hline
Discr. & \multicolumn{6}{c||}{$k=6$, $\Delta t=1$} & \multicolumn{6}{c|}{$k=6$, $\Delta t=0.1$}\\\hline
Solver & \multicolumn{3}{c|}{BILU-DG} & \multicolumn{3}{c||}{MG$_{\textrm{full}}$-DG} & \multicolumn{3}{c|}{BILU-DG} & \multicolumn{3}{c|}{MG$_{\textrm{full}}$-DG}\\\hline
$n_p$ & Time & $E$ & $IT_a$ & $SU_{MB}$ & $E$ & $IT_a$ & Time & $E$ & $IT_a$ & $SU_{MB}$ & $E$ & $IT_a$ \\
1 & 533.69 &  & 87.33 & 1.73 &  & 5.67 & 390.37 &  & 32.00 & 1.90 &  & 3.00 \\
2 & 381.18 & 0.70 & 190.83 & 2.37 & 0.96 & 5.83 & 247.56 & 0.79 & 80.83 & 2.29 & 0.95 & 3.00 \\
4 & 198.53 & 0.67 & 213.50 & 2.42 & 0.94 & 5.83 & 123.33 & 0.79 & 85.33 & 2.25 & 0.94 & 3.00 \\
8 & 109.00 & 0.61 & 257.50 & 2.69 & 0.95 & 5.33 & 62.84 & 0.78 & 99.17 & 2.24 & 0.92 & 3.00 \\
16 & 63.14 & 0.53 & 313.17 & 2.55 & 0.78 & 6.33 & 32.33 & 0.75 & 105.83 & 2.04 & 0.81 & 3.00 \\
32 & 47.91 & 0.35 & 392.33 & 2.65 & 0.53 & 7.17 & 19.61 & 0.62 & 110.67 & 1.86 & 0.61 & 3.00 \\\hline\hline
Solver & \multicolumn{3}{c|}{BILU-$m$HDG} & \multicolumn{3}{c||}{MG$_{\textrm{full}}$-$m$HDG} & \multicolumn{3}{c|}{BILU-$m$HDG} & \multicolumn{3}{c|}{MG$_{\textrm{full}}$-$m$HDG}\\\hline
$n_p$ & $SU_{MB}$ & $E$ & $IT_a$ & $SU_{MB}$ & $E$ & $IT_a$ & $SU_{MB}$ & $E$ & $IT_a$ & $SU_{MB}$ & $E$ & $IT_a$ \\
1 & 1.46 &  & 46.50 & 1.45 &  & 2.67 & 1.08 &  & 23.17 & 1.07 &  & 2.17 \\
2 & 1.75 & 0.84 & 112.50 & 1.76 & 0.85 & 2.67 & 1.16 & 0.85 & 50.67 & 1.16 & 0.85 & 2.17 \\
4 & 1.58 & 0.73 & 139.50 & 1.61 & 0.75 & 2.67 & 1.01 & 0.74 & 54.50 & 1.01 & 0.74 & 2.17 \\
8 & 1.58 & 0.66 & 155.83 & 1.61 & 0.68 & 2.83 & 0.94 & 0.68 & 62.17 & 0.93 & 0.68 & 2.17 \\
16 & 1.52 & 0.55 & 203.83 & 1.59 & 0.58 & 3.67 & 0.83 & 0.58 & 71.83 & 0.83 & 0.58 & 2.17 \\
32 & 1.74 & 0.42 & 268.50 & 1.87 & 0.45 & 5.17 & 0.80 & 0.46 & 82.50 & 0.80 & 0.46 & 2.17 \\\hline\hline
Solver & \multicolumn{3}{c|}{BILU-$p$HDG} & \multicolumn{3}{c||}{MG$_{\textrm{full}}$-$p$HDG} & \multicolumn{3}{c|}{BILU-$p$HDG} & \multicolumn{3}{c|}{MG$_{\textrm{full}}$-$p$HDG}\\\hline
$n_p$ & $SU_{MB}$ & $E$ & $IT_a$ & $SU_{MB}$ & $E$ & $IT_a$ & $SU_{MB}$ & $E$ & $IT_a$ & $SU_{MB}$ & $E$ & $IT_a$  \\
1 & 3.15 &  & 45.83 & 3.11 &  & 2.50 & 2.36 &  & 22.17 & 2.30 &  & 2.00 \\
2 & 3.71 & 0.83 & 106.50 & 3.81 & 0.86 & 2.50 & 2.52 & 0.84 & 48.50 & 2.50 & 0.86 & 2.00 \\
4 & 3.35 & 0.71 & 133.00 & 3.46 & 0.75 & 2.67 & 2.20 & 0.74 & 52.50 & 2.18 & 0.75 & 2.00 \\
8 & 3.35 & 0.65 & 149.33 & 3.46 & 0.68 & 2.67 & 2.04 & 0.67 & 59.50 & 2.03 & 0.69 & 2.00 \\
16 & 3.19 & 0.54 & 193.33 & 3.39 & 0.58 & 3.50 & 1.80 & 0.58 & 69.17 & 1.78 & 0.58 & 2.00 \\
32 & 3.38 & 0.37 & 257.00 & 3.84 & 0.43 & 5.17 & 1.73 & 0.46 & 79.67 & 1.71 & 0.46 & 2.00 \\\hline
\end{tabular}
\caption{Computational efficiency comparison using DG and HDG discretizations. Laminar vortex test case at $Re=100$, $M=0.05$, discretized using $16{\times}16$ mesh with $\mathbb{P}_6$ polynomials. Two time steps, $\Delta t=\{1;0.1\}$ using ESDIRK3 are reported. $SU_{MB}$ stands for the speed-up factor referred to the DG, matrix-based, BILU-preconditioned computation, $E$ is the parallel efficiency and $IT_a$ the average number of GMRES iterations.}
\label{tab:P6ViscVortPar}
\end{table}
Table~\ref{tab:P3ViscVortPar} compares the performance of the multigrid preconditioner to those of BILU for $k=3$ using $\Delta t=1$ (left) and $\Delta t=0.1$ (right), for three space discretizations, \emph{i.e.} DG, $m$HDG and $p$HDG. For all the three space discretization, the performance degradation of the BILU preconditioner shows up clearly in view of the considerable increase in the number of GMRES iterations moving from serial to parallel runs. On the other hand, the multigrid preconditioning strategy shows higher parallel efficiencies in all cases, with the increase in number of iterations either very low (for the largest time step) or non-existent (for the smallest time step). In fact, with the exception of the coarsest space solution, the algorithm is not affected by the domain decomposition.

Regarding the CPU time, we report the speed-up values non-dimensionalized by the CPU time of the DG, matrix-based and BILU-preconditioned computation. For the largest time step size, switching from single-grid BILU to multigrid provides consistent benefits on all the three space discretizations in view of a higher parallel efficiency of the approach. In fact, despite being $SU_{MB}$ lower for serial computations, it reaches its peak for 32 cores. The strategy provides a speedup of 2.09 for DG, 2.21 and 3.05 for $m$HDG and $p$HDG, respectively. 

For the smaller time step the situation is slightly different. In fact, the reduced conditioning of the matrix reduces the iterative solution times, as well as increases the relative cost of the matrix assembly. For $m$HDG, where the assembly costs are the highest, we report speed-up values below one. Conversely, $p$HDG still outperforms the reference providing a speed-up factor in the range $[1.29,1.75]$ due to the smaller number of operations during the Jacobian assembly with respect to the \emph{mixed} form. However, as opposed to what happens for DG, where an improvement in computational efficiency is still observed, the use of a $p$-multigrid does not benefit the performance of the solver.

Table~\ref{tab:P6ViscVortPar} shows the same results for a $k=6$ space discretization. Similar observations to those reported for Table~\ref{tab:P3ViscVortPar} can be made on the overall parallel behaviour of the preconditioning strategies here considered, \emph{i.e.} the increase in the number of iterations of the preconditioner reflects the performance degradation of the solution strategy when applied in parallel. However, in this case, the advantages arising from the use of a multigrid preconditioning strategy are more evident. In fact, for the largest time step size, the speed-up values are higher than those reported in Table~\ref{tab:P3ViscVortPar} involving 3\textsuperscript{rd} order polynomials. In particular, when using 32 cores, the speedup reaches 2.65, 1.87 and 3.84 for DG, $m$HDG and $p$HDG respectively. Differently to what was observed previously, when reducing the time step size, the advantages of using a multigrid strategy still appear evident for DG, which provides speed-ups in the range $[1.86,2.29]$. While for higher-order polynomials and smaller time steps a \emph{mixed} HDG implementation provides performance in line to that of a matrix-based, BILU-DG solver, the \emph{primal} implementation shows to be better performing overall, even if it shows an overall lower parallel efficiency. In particular, BILU-$p$HDG is the best performing solution strategy providing speed-up values in the range $[1.73,2.36]$. In this case, $p$-multigrid performs similarly to BILU. 

\begin{table}[t!]
\footnotesize
\centering
\begin{tabular}{|c||c||cc|cc||cc|cc|}
\hline
 & &\multicolumn{4}{c||}{$\Delta t = 1$} & \multicolumn{4}{c|}{$\Delta t = 1/10$} \\\hline
 & &\multicolumn{2}{c|}{BILU-DG} & \multicolumn{2}{c||}{MG$_{\textrm{full}}$-DG} & \multicolumn{2}{c|}{BILU-DG}  & \multicolumn{2}{c|}{MG$_{\textrm{full}}$-DG} \\\hline
Case & $n_p$ &MF & MFL & MF & MFL & MF & MFL & MF & MFL \\\hline\hline
\multirow{6}{*}{$k=3$} & 1 & 0.73 & 0.97 & 0.76 & 0.89 & 0.84 & 1.53 & 0.67 & 0.85 \\
 & 2 & 0.68 & 0.80 & 1.02 & 1.21 & 0.74 & 1.09 & 0.81 & 1.04 \\
 & 4 & 0.65 & 0.72 & 1.07 & 1.28 & 0.72 & 1.03 & 0.78 & 1.02 \\
 & 8 & 0.61 & 0.73 & 1.13 & 1.32 & 0.68 & 0.91 & 0.80 & 1.03 \\
 & 16 & 0.56 & 0.63 & 1.20 & 1.36 & 0.61 & 0.79 & 0.72 & 0.90 \\
 & 32 & 0.59 & 0.62 & 1.41 & 1.59 & 0.58 & 0.71 & 0.69 & 0.86\\ \hline\hline
\multirow{6}{*}{$k=6$} & 1 & 1.03 & 2.10 & 1.84 & 2.56 & 1.00 & 3.10 & 1.97 & 3.36 \\
 & 2 & 0.99 & 1.47 & 2.42 & 3.18 & 0.99 & 2.03 & 2.33 & 3.98 \\
 & 4 & 0.96 & 1.37 & 2.41 & 3.32 & 0.98 & 1.92 & 2.23 & 3.78 \\
 & 8 & 0.93 & 1.25 & 2.69 & 3.78 & 0.95 & 1.75 & 2.18 & 3.65 \\
 & 16 & 0.85 & 1.06 & 2.35 & 3.11 & 0.90 & 1.56 & 1.90 & 3.15 \\
 & 32 & 0.78 & 0.98 & 2.37 & 3.03 & 0.86 & 1.41 & 1.70 & 2.77 \\\hline
\end{tabular}
\caption{Computational efficiency of a matrix-free DG strategy. Laminar vortex test case at $Re=100$, $M=0.05$, discretized using $16{\times}16$ mesh with $\mathbb{P}_3$ and $\mathbb{P}_6$ polynomials. Two time steps, $\Delta t=\{1;0.1\}$ using the ESDIRK3 scheme are reported. $SU_{MB}$ stands for the speed-up factor referred to the matrix-based, BILU-DG computation reported in Tables~\ref{tab:P3ViscVortPar} and~\ref{tab:P6ViscVortPar}, $SU_{MF}$ is the speedup computed considering the matrix-free BILU-DG settings. Results obtained by lagging the preconditioner evaluation (MFL) for the entire solution process, \emph{i.e.} six linear systems, are also reported.}
\label{tab:DGmatfree}
\end{table}

Table~\ref{tab:DGmatfree} reports for comparison the speed-up values obtained using a matrix-free implementation of the iterative solver for the DG space discretization. In particular, we compute the matrix-based speedup $SU_{MB}$ using as a reference the matrix-based, BILU-DG solver to show the performance compared to the reference algorithm. The performance are evaluated using the same preconditioners reported in Tables~\ref{tab:P3ViscVortPar} and~\ref{tab:P6ViscVortPar}.
Considering the single-grid matrix-free, BILU-DG numerical experiments, one can summarize that:
\begin{enumerate}
    \item For $k=3$, switching MB to MF penalizes the CPU time. The reason for this is two-fold: first, the serial computation is 35\% slower than the MB one. Second, when the solver is applied in parallel, the matrix-free implementation seems to be less parallel efficient, since the speed-up factor decreases by increasing the number of processors;    
    \item for $k=6$, the penalization decreases. In fact, in serial computation, the same computational time has been recorded, meaning that a matrix-free iteration performs similarly to a matrix-vector product. However, a lower parallel efficiency is still observed.
    \item When a small time step is employed, higher speed-up values are achieved if compared to MB. In other words, the lower the number of GMRES iterations, the higher the performance. A slightly higher parallel efficiency is also observed.
\end{enumerate}

The possibility of lagging the preconditioner evaluation is also explored and the results are reported in Table~\ref{tab:DGmatfree} (MFL). In particular, we skip the recomputation of the preconditioner for six consecutive iterations, which means that the Jacobian matrix is evaluated only for the first non-linear iteration of the first stage of the ESDIRK3 scheme. By doing so, it is observed that the linear system converges with the same number of iterations which are not reported for brevity. This shows, at least for this kind of problem, that the preconditioner does not loose its efficiency throughout the stages of the same time step. Moreover, it confirms the powerful properties of the matrix-free iterative strategy, which allow skipping the Jacobian evaluation without degrading the convergence of Newton's method.

From the CPU time point of view, lagging the preconditioner improves the computational efficiency, since the matrix is evaluated only once. This is reflected by the speed-up values (see the MFL result columns). We point out that the maximum speed-up values are obtained for high orders, when the impact of the Jacobian assembly is large on the overall cpu time, and for small time step sizes: in this case the conditioning of the matrix and the CPU time spent on the iterative solution process reduce, and so the computational time saved by skipping the Jacobian assembly reflects on the overall efficiency of the method. It is worth pointing out that by doing so the matrix-free penalization at low orders is reduced, while speed-up values in the range $[1.41,3.10]$ are achieved for the $k=6$, $\Delta t=0.1$ case.

Similar observations hold true when the matrix-free approximation is employed within a multigrid strategy. Also in this case the problem has been solved using the same number of GMRES iterations with respect to the non-lagging computation. However, from the CPU time of view, a higher speed-up value is achieved both in serial and in parallel runs thanks to the optimal multigrid scalability with respect to single-grid BILU preconditioning. The speed-up factor for the largest polynomial order is now in the range $[2.56,3.78]$ and $[2.77,3.98]$ for the large and small time step, respectively. Those speed-up values are larger than the one obtained for a fully matrix-based implementation of the DG discretization, see Table~\ref{tab:P6ViscVortPar}. In comparison to HDG, it can be seen that by using the Jacobian lagging option, larger speed-up values are generally achieved for small time step sizes at high order. However, the use of MG$_{\textrm{full}}$-$p$HDG may be preferred to MG$_{\textrm{full}}$-DG, even coupled with matrix-free and preconditioner lagging for the largest time step employed, as a better suitability to deal with very stiff and high-order discretizations have been highlighted. 

We remark that, for computational efficiency, we employ the matrix-free implementation of the iterative solver only on the finest space of the multilevel iterative solution, similarly to what has been done in~\cite{franciolini2018p}. This choice is consistent with the results obtained for the single-grid preconditioner in Table~\ref{tab:DGmatfree}. In fact, the matrix-free iteration cost is similar to that of a matrix-based one only for high orders, while it is larger for lower-order polynomials.  It therefore is appropriate to employ matrix-based smoothers on the coarse levels. When discretizing the equations using high orders, this idea seems to work optimally. Note also that the overall memory footprint of the application is dominated by the allocation of the block-Jacobi preconditioner for the finest space smoother, as shown in Figure~\ref{fig:MemTot}.


\subsection{Remarks}


We here summarize the main points of the previous section. First, we see clearly that having large time step sizes maximizes the advantages of coupling HDG and $p$-multigrid strategy, since the expenses of using static condensation together with a more powerful and expensive preconditioning strategy produce a faster solver only for high conditioning of the system, while the higher scalability of the multigrid algorithm as opposed to single-grid reflect on the overall scalability of the solver. On the other hand, for small time step sizes, the computational time is dominated by the Jacobian assembly and condensation for HDG, thus the CPU time as well as the parallel efficiency is dominated by the matrix assembly.

For HDG, we demonstrate how the \emph{mixed} and \emph{primal} formulations provide comparable results in terms of error levels by refining both in space and time, despite the fact that the \emph{primal} form provides a one order lower convergence rate for the gradient variable. From the algorithmic point of view, the statically condensed system is typically solved using roughly the same number of iterations. However, the computational time is considerably lower for the $p$HDG formulation, since it does not deal with the Jacobian entries related to the state gradient. For this reason, in the remainder of the work, only the \emph{primal} formulation will be employed for benchmarking. We stress that even for small time step sizes, the single-grid preconditioned $p$HDG solver performs fairly well in comparison to the other solution strategies, showing the benefits of the approach for unsteady flow computations.

We remark that the current implementation of the parallelization strategy makes HDG less parallel efficient than DG. This fact is ascribed to the higher amount of duplicate operations due to the integration over halo mesh elements and faces created by the domain decomposition required for the static condensation of the interior degrees of freedom of the Jacobian matrix. On the other hand, in DG the duplicate work involves only partition faces. Other implementation choices, such as those related to the minimization of the duplicate work on the partition boundaries, may be considered in future works.

We finally point out that, despite being appealing, a matrix-free implementation of the hybridizable discontinuous Galerkin method is not at all straightforward for obtaining satisfactory performance in CPU time~\cite{kronbichler2016performance}, and thus its development is beyond the scope of the present work.

\section{Results on complex test cases}
The second family of numerical experiments deals with the solution of two test cases: i) laminar flow over a two-dimensional circular cylinder at $Re=100$ and $M=0.2$~\cite{meneghini2001numerical}; and ii) the solution of the plunging motion of a NACA 0012 airfoil at $Re=1000$ and Mach number $M=0.2$. The latter case is solved by using the ALE mesh motion formulation introduced in~\cite{fidkowski2016hybridized}. Those results are devoted to extend the comparison to more complex unsteady flow problems. In all the cases reported herein, the right-preconditioning approach is still employed such that the convergence of the linear solver is not affected by changing the preconditioning operator, and therefore all the numerical experiments are performed using a similar accuracy.

\subsection{Circular cylinder}

\begin{figure}[t!]
\centering
\includegraphics[trim={0 7cm 0 7cm},clip=true,width=11cm]{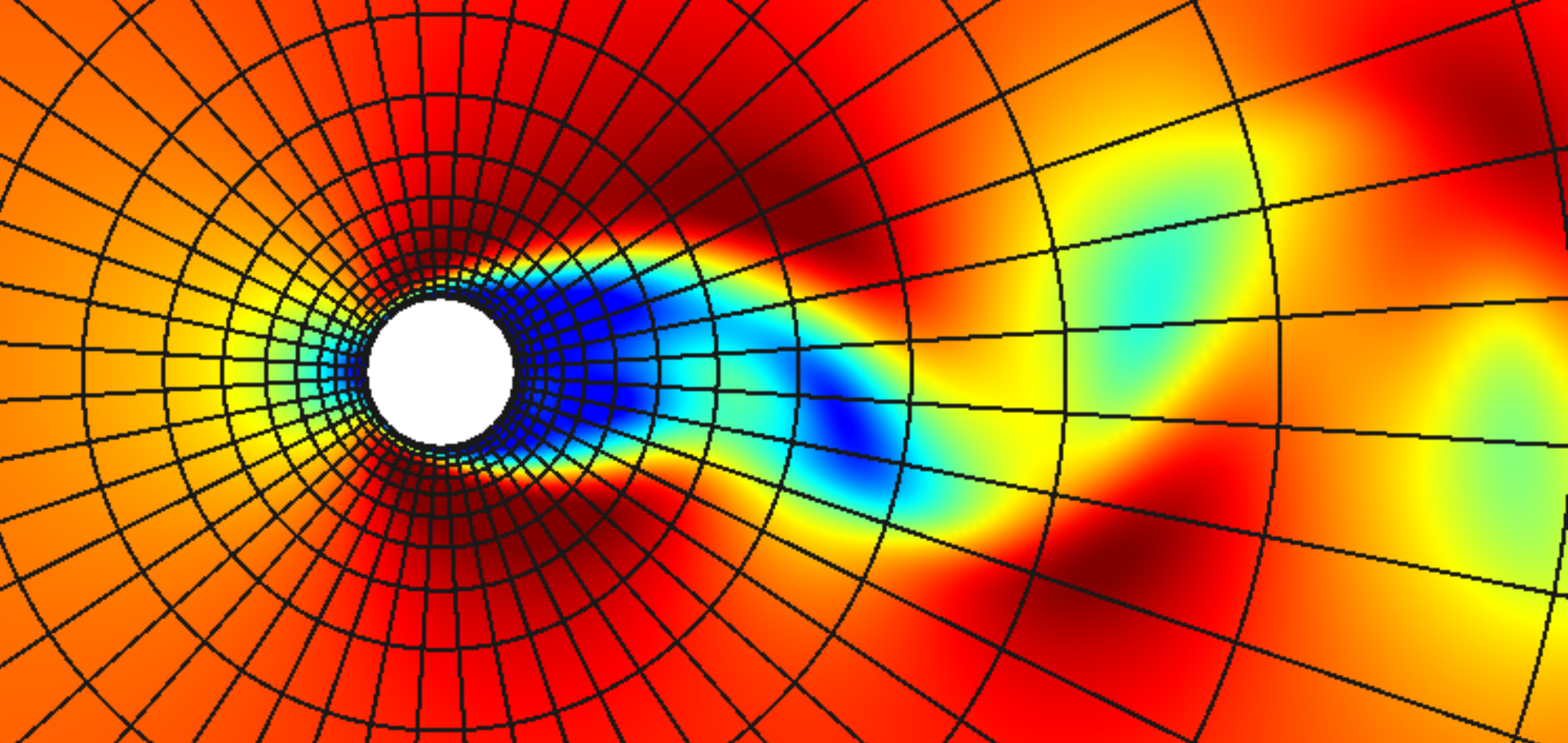}
\caption{Laminar flow around a circular cylinder at $Re=100$, $M=0.2$. Mach number contours.}
\label{fig:cyl2d}
\end{figure}

Laminar flow around a circular cylinder at Mach number $M=0.2$ and Reynolds number $Re=100$ has been solved on a grid of $n_e=960$ mesh elements using a $\mathbb{P}_6$ space discretization. Figure~\ref{fig:cyl2d} shows a snapshot of the computed Mach number contours. To integrate the governing equations in time, the four-stage, third-order explicit-first-stage, diagonally-implicit Runge--Kutta method (ESDIRK3) was employed. The solution accuracy was assessed by comparing with literature data~\cite{meneghini2001numerical}. To this end, Table~\ref{tab:CYL_COEFF} shows the averaged drag and lift coefficients, as well as the Strouhal number of the body forces ($C_d$, $C_l$, $St$) for several temporal refinements on the same grid. The coefficients were obtained by averaging a statistically-developed solution over ten shedding periods. An overall good agreement has been found, while a temporal convergence can be observed by using a non-dimensional time step of $\Delta t \leq 0.25$. It is well known~\cite{franciolini2017efficiency} that the time step size greatly affects the iterative solution process, and therefore it has to be taken into account to evaluate the performance. Presently, we use the largest time step size that yields both converged body forces and Strouhal numbers, and this maximizes the efficiency of the solution strategy. Considering the results in Table~\ref{tab:CYL_COEFF}, we choose $\Delta t=0.25$.
\begin{table}[t!]
\centering
\begin{tabular}{|l||l|l|l|}
\hline
$(U/L)\Delta t$ & $C_d$ & $C_l$ & $St$\\\hline\hline
0.5 & 1.3468 & 3.383e-03 & 0.16327\\
0.25 & 1.3519 & -1.441e-03 & 0.16410\\
0.125 & 1.3527 & -1.400e-04 & 0.16410\\
0.05 & 1.3528 & -1.718e-04 & 0.16410\\ 
0.025 & 1.3528 &-6.353e-06 & 0.16410\\\hline
\end{tabular}
\caption{Laminar vortex test case at $Re=100$, $M=0.05$. Time convergence rates for the DG and HDG spatial discretizations and the ESDIRK3 temporal scheme.}
\label{tab:CYL_COEFF}
\end{table}
The parallel performance of the solution strategies introduced in the previous sections is assessed by considering the effects of domain decomposition. To do so, a fully-developed flow field is integrated in time for 10 time steps to compute the average number of GMRES iterations and the convergence rates during the non-linear solution. The computations are performed on the range of 1 to 64 cores ($n_p$) on a platform based on two 16-core AMD Opteron processors arranged in a two-processor per-node fashion, for a total of 32 cores per node. A fixed relative tolerance of $10^{-6}$ to stop the GMRES solver is used, as well as an absolute tolerance of $10^{-5}$ for the Newton-Raphson method.

\begin{table}[t!]
\footnotesize
\centering
\begin{tabular}{|c||ccc|c|c|ccc|}
\hline
Preconditioner & \multicolumn{8}{c|}{BILU} \\ \hline
Case & \multicolumn{3}{c|}{DG-MB} & \multicolumn{1}{c|}{DG-MF} & \multicolumn{1}{c|}{DG-MFL} & \multicolumn{3}{c|}{$p$HDG} \\\hline\hline
$n_p$ & Time & $E$ & $IT_a$ & $SU_{MB}$ & $SU_{MB}$ & $SU_{MB}$ & $E$ & $IT_a$ \\\hline
1 & 17425.77 &  & 19.43 & 1.00 & 3.90 & 2.18 &  & 17.40 \\
2 & 11321.70 & 0.77 & 60.65 & 1.00 & 1.99 & 2.68 & 0.94 & 29.14 \\
4 & 5907.54 & 0.74 & 70.39 & 0.99 & 1.67 & 2.65 & 0.89 & 34.16 \\
8 & 2948.54 & 0.74 & 70.15 & 0.98 & 1.65 & 2.50 & 0.84 & 35.30 \\
16 & 1551.90 & 0.70 & 85.41 & 0.94 & 1.49 & 2.40 & 0.77 & 39.15 \\
32 & 815.46 & 0.67 & 98.24 & 0.90 & 1.35 & 2.19 & 0.67 & 45.11 \\
64 & 692.93 & 0.39 & 117.33 & 0.86 & 1.19 & 2.09 & 0.38 & 50.53 \\\hline
\hline
Preconditioner & \multicolumn{8}{c|}{MG$_{\textrm{full}}$} \\ \hline
Case & \multicolumn{3}{c|}{DG-MB} & \multicolumn{1}{c|}{DG-MF} & \multicolumn{1}{c|}{DG-MFL} & \multicolumn{3}{c|}{$p$HDG} \\\hline\hline
$n_p$ & $SU_{MB}$ & $E$ & $IT_a$ & $SU_{MB}$ & $SU_{MB}$ & $SU_{MB}$ & $E$ & $IT_a$ \\\hline
1 & 1.70 &  & 4.00 & 1.75 & 2.24 & 2.13 &  & 3.03 \\
2 & 2.12 & 0.96 & 4.00 & 2.17 & 2.77 & 2.62 & 0.95 & 3.03 \\
4 & 2.20 & 0.95 & 4.00 & 2.20 & 3.15 & 2.58 & 0.89 & 3.03 \\
8 & 2.14 & 0.93 & 4.00 & 2.10 & 2.96 & 2.45 & 0.85 & 3.03 \\
16 & 2.15 & 0.89 & 4.00 & 2.04 & 2.92 & 2.35 & 0.77 & 3.03 \\
32 & 2.12 & 0.83 & 4.00 & 1.91 & 2.69 & 2.14 & 0.67 & 3.03 \\
64 & 2.07 & 0.48 & 4.00 & 1.76 & 2.43 & 2.00 & 0.37 & 3.03 \\\hline
\end{tabular}
\caption{Circular cylinder test case at $Re=100$, $M=0.2$, discretized using $960$ mesh elements with $\mathbb{P}_6$ polynomials. Computational efficiency comparison of DG and $p$HDG solvers using BILU and $p$-multigrid, respectively. $SU_{MB}$ stands for the speed-up factor referred to the DG[GMRES(BILU);MB] computation, $E$ is the parallel efficiency and $IT_a$ the average number of GMRES iterations.}
\label{tab:2DCylComparison}
\end{table}
Table~\ref{tab:2DCylComparison} report the results of the computations. As observed for the convected vortex test case, the BILU preconditioner shows an increase in the number of GMRES iterations with increasing number of processes in both the DG and $p$HDG space discretizations. This behavior is attributed to the way the incomplete lower-upper factorization is performed, as it deals with the square, partition-wise block of the iteration matrix. Therefore, the preconditioner effectiveness naturally decreases as $n_p$ grows, since the number of off-diagonal blocks neglected by the ILU increases with $n_p$. By switching from MB to MF, the number of iterations remains the same and it is not reported for brevity. On serial computations, the CPU time remains in the same line since a speed-up value (SU$_{MB}$) of about one is obtained. When the solver is applied in parallel, a slightly lower parallel efficiency is observed, as the speed-up value decreases by increasing $n_p$. The possibility of lagging the Jacobian evaluation during the nonlinear solution process, as well as within a single time step through multiple stages of the scheme, is also explored and the results are reported in the column labelled as MFL. In this case a large speed-up value is achieved especially for serial computations, where ILU(0) works remarkably well. On the other hand, when this solution strategy is applied in parallel, a loss in parallel efficiency is observed and the solver performs in line with the reference, matrix-based method. 

Considering $p$HDG with BILU preconditioning, the iterative solution process still suffers of algorithmic degradation by parallelizing the computation, despite the increase in the number of iterations being smaller than that observed for DG. Nevertheless, the solver appears to be faster, with the speed-up value in the range [2.09, 2.68]. The parallel efficiency is also higher than that of the referece DG method: $p$HDG involves expensive local-to-each element operations that affect the computational time of the solver, but they scale ideally on multicore systems. In addition, a lower $IT_a$ is generally observed when employing this space discretization.

On the lower part of the table, numerical experiments performed using multigrid preconditioning are reported. Considering the matrix-based DG solver, it is possible to achieve an ideal parallel efficiency of the solver in terms of number of iterations by using multigrid, since the number of iterations remains constant at 4.00 in each case. The speed-up in this case is in the range [1.7, 2.2], which is consistent with what has been reported previously. By using a matrix-free implementation of the finest-level smoother and linear solver, similar performance for the non-lagged preconditioner is observed. As opposed to this, lagging the multigrid linear solver operator produces a speed-up value in the range [2.24, 3.15], which suggests this strategy to be the best performing of all the numerical experiments. In fact, the use of multigrid preconditioning in $p$HDG does not improve the computational efficiency of the solver over BILU-$p$HDG, despite reducing considerably the number of iterations and providing optimal algorithmic scalability. Observing the $p$HDG results, one can understand that with such time step size, the costs of statically condensing out the interior DoFs, as well as the back-solve to recover them from the trace DoFs, dominate the percentage of the overall computational time. It is worth mentioning that in such a practical application the $p$-multigrid preconditioning approach for HDG works very well from the number of iterations point of view, and that the same numerical set-up as that of a DG solver reported in Table~\ref{tab:MGSettings} has been employed to precondition the system.

As a final comparison, Table~\ref{tab:2DCylTri} reports the same test case solved using a grid obtained by splitting the quadrilateral elements into triangles. Full-order basis functions were employed. Only the computation with $n_p=64$ is reported. It is worth pointing out that this space discretization reduces the total number of DoFs by the 23.8\% with respect to the previous one. By comparing the computational time and average number of GMRES iterations, similar speed-up values are obtained using the DG-MB solver as well as $p$HDG. On the other hand, the DG-MF solver seems to be slightly penalized with respect to the MB one, as the speed-up values of the computations drop by a factor between 7 to 15 percent. It is worth pointing out that the matrix-free penalization that arise in this case is in line to what reported in previous studies~\cite{franciolini2017efficiency,franciolini2018p} using broken polynomial spaces, which reduce by increasing the number of DoFs per element, for example in three-dimensional computations. On the other hand, a slightly larger speedup for $p$HDG computations has been observed, which makes this solution strategy the most convenient from the CPU time point of view.

\begin{table}[t!]
\footnotesize
\centering
\begin{tabular}{|c||cc|c|c|cc|}
\hline
Preconditioner & \multicolumn{6}{c|}{BILU} \\ \hline
Case & \multicolumn{2}{c|}{DG-MB} & \multicolumn{1}{c|}{DG-MF} & \multicolumn{1}{c|}{DG-MFL} & \multicolumn{2}{c|}{$p$HDG} \\\hline\hline
$n_p$ & Time & $IT_a$ & $SU_{MB}$ & $SU_{MB}$ & $SU_{MB}$ & $IT_a$ \\\hline
64 & 322.26 & 129.050 & 0.80 & 1.00 & 2.40 & 49.750 \\\hline
\hline
Preconditioner & \multicolumn{6}{c|}{MG$_{\textrm{full}}$} \\ \hline
Case & \multicolumn{2}{c|}{DG-MB} & \multicolumn{1}{c|}{DG-MF} & \multicolumn{1}{c|}{DG-MFL} & \multicolumn{2}{c|}{$p$HDG} \\\hline\hline
$n_p$ & $SU_{MB}$ & $IT_a$ & $SU_{MB}$ & $SU_{MB}$ & $SU_{MB}$ & $IT_a$ \\\hline
64 & 1.66 & 4.525 & 1.43 & 1.77 & 2.05 & 3.388 \\\hline
\end{tabular}
\caption{Circular cylinder test case at $Re=100$, $M=0.2$, discretized using $1920$ mesh elements with $\mathbb{P}_6$ full-order basis functions. Computational efficiency comparison of DG and $p$HDG solvers using BILU and $p$-multigrid, respectively. $SU_{MB}$ stands for the speed-up factor referred to the matrix-based, BILU-DG computation and $IT_a$ for the average number of GMRES iterations.}
\label{tab:2DCylTri}
\end{table}

\subsection{Laminar flow around a heaving and pitching NACA 0012 airfoil}
\begin{figure}[htbp!]
\centering
\includegraphics[trim={0 7cm 0 7cm},clip=true,width=11cm]{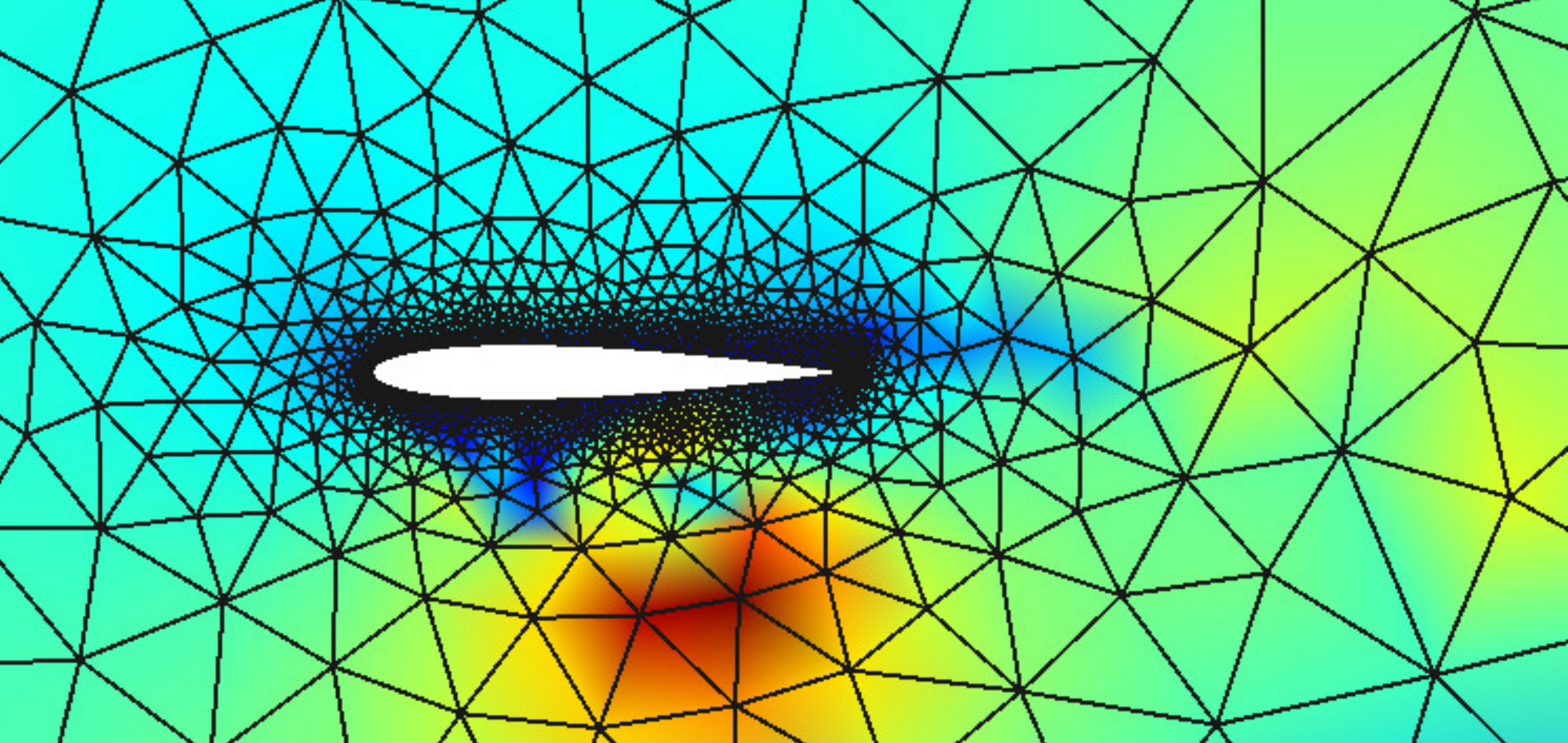}
\caption{Laminar flow around a heaving NACA 0012 airfoil. $Re=1000$, $M=0.2$. Mach number contours at the final time $T=2$.}
\label{fig:naca}
\end{figure}
This test case is the CL1 (heaving and pitching airfoil) proposed for the 5\textsuperscript{th} international workshop on high-order CFD methods (HiOCFD5), see~\cite{5HHW}. The simulation involves the compressible Navier--Stokes equations, with $\gamma=1.4$, $Pr=0.72$, and constant viscosity, to simulate flow over a moving NACA 0012 airfoil. The airfoil is modified to obtain a closed trailing edge via the equation
\begin{equation}
y(x) = \pm 0.6\left(0.2969\sqrt{x} - 0.1260x - 0.3516x^2 +0.2843x^3 - 0.1036x^4\right),
\end{equation}
with $x\in [0,1]$. The initial condition is a steady state solution at a free-stream Mach number of $M=0.2$ and a Reynolds number $Re =1000$. The motion is a pure plunging motion governed by the following function
\begin{equation}\label{eq:plunge}
    h(t) = t^2(3-t)/4,
\end{equation}
and the output of interest is the total energy and vertical impulse exchanged between the airfoil and the fluid,
\begin{equation}
    W=\int_{T}{F_y(t)\dot{h}(t)dt}, \qquad \qquad I=\int_{T}{F_y(t)dt},
\end{equation}
where $F_y(t)$ is the vertical force computed on the airfoil surface, and $\dot{h}(t)$ is the time derivative of Eq.~\eqref{eq:plunge}. In particular, the output is evaluated at a non-dimensional time $T=2$. The mesh consists of a triangulation of the domain by $n_e=2137$ elements, shown in Figure~\ref{fig:naca}, and the solution approximation space is $\Pol_6$.

The time step size of the simulation has been evaluated through a time-convergence analysis of the output quantities, reported in Table~\ref{tab:Naca0012Results}.
\begin{table}[t!]
\footnotesize
\centering
\begin{tabular}{|c|c|c|}
\hline
$T/\Delta t$ & $W$ & $I$ \\\hline\hline
20 & -1.3860 & -2.3667 \\\hline
40 & -1.3839 & -2.3444 \\\hline
80 & -1.3837 & -2.3391 \\\hline
160 & -1.3837 & -2.3378 \\\hline
\end{tabular}
\caption{Time convergence analysis of the laminar flow around a plunging NACA 0012 airfoil. A satisfactory accuracy is observed for $T/\Delta t=2$, where $T=2$ is the length of the simulation and $\Delta t$ is the time step size.}
\label{tab:Naca0012Results}
\end{table}
\begin{table}[t!]
\footnotesize
\centering
\begin{tabular}{|c||cc|c|c|cc|}
\hline
Preconditioner & \multicolumn{6}{c|}{BILU} \\ \hline
Case & \multicolumn{2}{c|}{DG-MB} & \multicolumn{1}{c|}{DG-MF} & \multicolumn{1}{c|}{DG-MFL} & \multicolumn{2}{c|}{$p$HDG} \\\hline\hline
$n_p$ & Time & $IT_a$ & $SU_{MB}$ & $SU_{MB}$ & $SU_{MB}$ & $IT_a$ \\\hline
64 & 900.19 & 136.119 & 0.69 & 0.81 & 2.35 & 49.750 \\\hline
\hline
Preconditioner & \multicolumn{6}{c|}{MG$_{\textrm{full}}$} \\ \hline
Case & \multicolumn{2}{c|}{DG-MB} & \multicolumn{1}{c|}{DG-MF} & \multicolumn{1}{c|}{DG-MFL} & \multicolumn{2}{c|}{$p$HDG} \\\hline\hline
$n_p$ & $SU_{MB}$ & $IT_a$ & $SU_{MB}$ & $SU_{MB}$ & $SU_{MB}$ & $IT_a$ \\\hline
64 & 1.49 & 4.477 & 1.06 & 1.13 & 2.13 & 2.754 \\\hline
\end{tabular}
\caption{Circular cylinder test case at $Re=100$, $M=0.2$, discretized using $2137$ mesh elements with $\mathbb{P}_6$ full-order basis functions. Computational efficiency comparison of DG and $p$HDG solvers using BILU and $p$-multigrid, respectively. $SU_{MB}$ stands for the speed-up factor referred to the matrix-based, BILU-DG computation and $IT_a$ for the average number of GMRES iterations.}
\label{tab:2DNacaLaminar}
\end{table}
Table~\ref{tab:2DNacaLaminar} compares the computational efficiency of the solution strategies in terms of the speed-up factor, $SU_{MB}$, and the number of iterations, $IT_a$. Only the computation with the larger number of cores is reported to show the efficiency on large and practical computations. The reference to compute the speed-up is the computational time of the matrix-based, BILU-DG solver. By switching from a matrix-based to a matrix-free implementation, the computational strategy is penalized by higher computational cost of a single iterations, see the MF column. By employing the Jacobian lagging (MFL), this penalization is reduced only slightly. On the other hand, the $p$HDG implementation provides a solver which is more than twice as much as fast, while the system require a considerably lower number of GMRES iterations with respect to the reference. The use of $p$-multigrid preconditinoing in a DG context provides a speed-up factor of about 1.49 for a fully matrix-based implementation, and the number of iterations drops from 136 to around 4.4 on average. The use of matrix-free in this case still penalizes the solver, which is about 12\% faster for the MFL case. Multigrid preconditioning is shown to be robust enough to precondition the linear system arising from the $p$HDG discretization, since it converges using an average of 2.754 iterations. However, this gain is not reflected on the CPU time, which is slightly higher.


\section{Conclusions}
The paper compares, within the same framework, the computational efficiency of different high-order discontinuous Galerkin implementations. The first involves a modal discontinuous Galekin method coupled with matrix-based and matrix-free iterative solvers, while the second one considers a hybridizable discontinuous Galerkin implementation, both in the \emph{mixed} form, which allocates the components of the Jacobians related to the gradient variable, and in the \emph{primal} form, which forgoes separate approximation of the gradient variable and symmetrizes the discretization by adding an additional adjoint-consistency term. The efficiency of the solution strategies is assessed by comparing different single-level preconditioners as well as multilevel ones such as $p$-multigrid, on a variety of two-dimensional test cases involving laminar viscous flows, including mesh motion, on meshes made by triangular and quadrilateral elements. The effects of parallel efficiency and the use of different basis functions have also been considered. The paper shows that the use of a matrix-free implementation of the iterative solver in the context of implicit discontinuous Galerkin discretizations provides a memory footprint which is in line to that of an HDG method if a block-diagonal preconditioner is employed within the smoother on the finest space. Moreover, the \emph{primal} HDG method becomes more efficient than the \emph{mixed} one, having a lower number of non-zeros Jacobian entries, and it provides comparable error levels. If compared to $p$HDG, the $p$-multigrid matrix-free solver is competitive in terms of CPU time when the problems involve time marching with small time steps, since the preconditioner evaluation can be lagged. On the other hand, HDG methods require expensive element-wise operations that become a bottleneck in those conditions. Finally, a novel approximate-inherited $p$-multigrid strategy has also been introduced for HDG. Such a strategy is more robust and efficient for different test cases and mesh types, being able to reduce considerably the number of iterations of the solution process. However, this gain in number of iterations is not reflected in the CPU time of the solver. Future work will be devoted to the validation of those strategies on stiff three-dimensional cases involving laminar and turbulent flows, and possibly hybrid RANS-LES models. 

\section*{Acknowledgements}
M. Franciolini acknowledge Dr. Lorenzo Botti from University of Bergamo for useful discussions and comments. Funding received from the Department of Energy under grant DE-FG02-13ER26146/DE-SC0010341 is also gratefully acknowledged.







\end{document}